\documentclass[preprint]{vgtc}               

\graphicspath{{figures/}{pictures/}{images/}{./}} 

\usepackage{times}                     

\usepackage{tabu}                      
\usepackage{booktabs}                  
\usepackage{lipsum}                    
\usepackage{mwe}                       
\usepackage{comment}

\usepackage{multirow}
\usepackage{subcaption}

\usepackage{mathptmx}                  

\usepackage{mathSymbols}
\usepackage{algorithm}
\usepackage{algpseudocode}
\usepackage{enumitem}

\newtheorem{lemma}[theorem]{Lemma}
\newtheorem{corollary}[theorem]{Corollary}

\newcommand{\real}[1]{
  \ifnum#1=1
    \mathbb{R}
  \else
    \mathbb{R}^{#1}
  \fi
}

\onlineid{0}

\vgtccategory{Research}

\vgtcinsertpkg

\preprinttext{To appear in an IEEE VGTC workshop ``TopoInVis Connect 2026 - Topology meets Artificial Intelligence''.}

\title{Exact Computation of Trait-induced Merge Trees for Bivariate Fields}

\author{
Petar Hristov\thanks{e-mail: petar.hristov@liu.se} 
\and Ingrid Hotz\thanks{e-mail: ingrid.hotz@liu.se}
\and Talha Bin Masood\thanks{e-mail: talha.bin.masood@liu.se}
}
\affiliation{\scriptsize Scientific Visualization Group, Department of Science and Technology (ITN), Link\"oping University, Norrk\"oping, Sweden}

\abstract{
Trait-induced merge trees (TIMTs) provide a robust topology-based method for selecting and browsing feature level sets in multivariate data by analyzing the distance field induced by a user-specified trait in attribute space. Existing TIMT computations typically sample this distance field at mesh vertices and assume piecewise-linear interpolation, although the Euclidean distance-to-trait function is generally not piecewise linear on the original mesh. As a result, the resulting merge tree may miss zero-valued features and may perturb the locations and values of minima and merge events. We study the exact computation of TIMTs for piecewise-linear bivariate fields, focusing first on point traits. We show that the restricted sublevel sets inside each tetrahedron are convex and therefore have trivial local merge-tree structure, implying that global topological changes arise only through gluing across simplex boundaries. Based on this observation, we construct a weighted graph whose merge tree is isomorphic to the exact merge tree of the induced distance field. 
We further relate TIMTs to Jacobi sets, showing how nonzero edge events of the TIMT are localized by the singular structure of the underlying bivariate map. We establish a theoretical upper bound on the error of the vertex-sampled linear interpolation, expressed in terms of the maximum length of projected mesh edges in the range. We discuss extensions to line, line-segment, and finite point-set traits, and implement the method robustly using CGAL and VTK, demonstrating results on both synthetic and real-world datasets.
} 

\keywords{Multifield topology, Merge trees, Traits, Features}

\begin{document}



\maketitle

\section{Introduction} 
\label{sec:introduction}

Simulations of natural phenomena often produce multiple interacting output variables, such as pressure, temperature, and precipitation in meteorology, density fields in computational chemistry, or tensor-derived quantities in mechanical engineering. Such multi-field data can be viewed as mappings from a spatial domain into a higher-dimensional attribute space. Although multi-fields are common in scientific applications, methods for their visualization and analysis remain less developed than corresponding methods for scalar or vector fields. This is particularly true for topological analysis. While scalar-field topology provides powerful descriptors for abstraction, simplification, and exploration~\cite{Heine2016TopologySTAR}, topological methods for multivariate fields remain comparatively difficult to apply in practice~\cite{Yan2021TopologySTAR}. Concepts such as Reeb spaces~\cite{Edelsbrunner2008ReebSpaces} and Jacobi sets~\cite{Edelsbrunner2004Jacobi} provide important theoretical foundations, but their practical use remains challenging.

Recent work has adapted scalar-field visualization techniques to multivariate data -- volume rendering using feature-space representatives~\cite{Jankowai2020TensorVolume}, fiber surfaces~\cite{Carr2015FiberSurfaces}, and feature level sets (FLS)~\cite{Jankowai2019FLS}. By defining a trait in attribute space that describes parameter configurations of interest, the corresponding distance field measures how closely each point in the domain matches this trait, and its level sets form the FLS. The zero level set coincides with the desired multivariate feature, such as a fiber surface in the bivariate case. Even if the zero level set is empty, users can still inspect nearby level sets to understand how closely the data approaches the specified trait.

Once features are defined by traits, one of the challenges is to select meaningful level values. Rendering many levels may create visual clutter, whereas using a single global value may overemphasize some regions and miss others. To address this challenge, Jankowai et al.~\cite{Jankowai2023TIMT,Lei2025TraitDesign} exploited merge tree-based feature selection~\cite{Weber2007TopologyVolRen} and introduced trait-induced merge trees (TIMTs), computed from the FLS distance field. The leaves of a TIMT correspond to regions closest to the trait and, therefore, to representative candidate features. The resulting hierarchy supports topology-based simplification, querying, and browsing of multivariate features. In particular, feature-selection strategies such as crown features~\cite{Nilsson2022Cyclones} can automatically identify relevant local isovalues and emphasize significant structures while reducing noise-induced clutter.

However, the exact computation of TIMTs has not been fully addressed. Jankowai et al.~\cite{Jankowai2023TIMT,Lei2025TraitDesign} acknowledge that the Euclidean distance-to-trait function is generally not piecewise linear, but compute TIMTs from distance values sampled at mesh vertices and then linearly interpolated over the domain. This approximation is useful for exploratory analysis, but it does not necessarily reflect the true evolution of sublevel sets of the distance transform. For example, even if the preimage of the trait is nonempty, the approximate TIMT need not contain a node of value zero. Similarly, the values and locations of the minima and merge events in the approximate TIMT may differ from those of the exact TIMT.

In this paper, we address the exact computation of TIMTs for piecewise-linear bivariate fields. We analyze the case of a single point trait in detail and provide an algorithm that constructs an exact merge tree via a graph whose sublevel-set connectivity matches that of the original distance field on the mesh. We implement the algorithm using VTK and CGAL and compare the exact TIMTs with the vertex-sampled approximation. We also establish theoretical bounds on the discrepancy between exact and approximate TIMTs, discuss extensions to more complex traits such as lines, line segments, and finite point sets, and outline connections between TIMTs and Jacobi sets. Our \textit{contributions} are as follows:
\begin{itemize}
    \item We formulate the exact computation problem for trait-induced merge trees of piecewise-linear bivariate fields and identify limitations of the vertex-sampled approximation.

    \item We prove structural properties of the point-trait distance field that lead to an exact construction of the TIMT. 

    \item We relate nonzero edge events of the TIMT to Jacobi edges of the underlying bivariate map, thereby connecting trait-induced topology with the singular structure of multifields.

    \item We establish an upper bound on the vertex-sampled interpolation error in terms of projected edge lengths in the range.

    \item We discuss extensions to more general multifields and to other traits such as line, line-segment, and finite point-set traits.
    
    \item We provide a robust CGAL/VTK implementation and demonstrate the method on synthetic and real-world datasets.
\end{itemize}

\section{Related Work}

\noindent\textbf{Inverse-image methods.} Several methods have been proposed for analysis of multifields using inverse image methods. Fiber surfaces extend isosurfaces to bivariate fields by extracting preimages of curves in the range, enabling richer feature characterization in multi-field settings~\cite{Carr2015FiberSurfaces}. Efficient and exact computation of fiber surfaces on tetrahedral meshes was later addressed by Klacansky et al.~\cite{Klacansky2017FiberSurfaces}, providing a practical foundation for their use in visualization pipelines. Subsequent work has extended these ideas to higher-dimensional and multi-variate settings, including fiber surfaces for many variables~\cite{Blecha2020FiberSurfaces} and feature level-sets~\cite{Jankowai2019FLS}, which generalize the concept of level sets to capture abstract features in multi-fields. These approaches highlight the importance of analyzing preimages of functions in the range space, a perspective that closely relates to the trait-based constructions considered in this work.

\vspace{0.5em}\noindent\textbf{Multifield topology.} To address the challenges of multi-field topology, several structures have been proposed to generalize scalar descriptors. Joint contour nets~\cite{Carr2014JCN} provide a discrete approximation of multi-field topology, while Reeb spaces~\cite{Edelsbrunner2008ReebSpaces} offer a continuous analogue capturing connectivity in the range space. Theoretical foundations for such constructions were established through the study of Jacobi sets~\cite{Edelsbrunner2004Jacobi} and singular fibers~\cite{Saeki2017SingularFibers}, which describe critical interactions between multiple scalar functions. More recent work has focused on efficient and correct algorithms for computing Reeb spaces for piecewise-linear bivariate fields, including arrange-and-traverse approaches~\cite{Hristov2025ATAlgorithm, Hristov2026SATAlgorithm} and robust geometric predicates~\cite{Hristov2025Predicates}. These contributions provide the algorithmic and theoretical backbone for understanding the topology of multi-field mappings.

Trait-induced merge trees (TIMTs) provide a recent and promising approach for analyzing multi-field data by reducing multi-variate structure to a scalar field analysis through the distance to trait-of-interest. The initial formulation of TIMTs was introduced by Jankowai et al.~\cite{Jankowai2023TIMT}, where traits encode application-specific features derived from multi-fields. This framework was further expanded to include systematic trait design and the integration of domain knowledge~\cite{Lei2025TraitDesign}, enabling flexible and interpretable multi-field visualization. By mapping complex multi-field behavior to scalar functions, TIMTs allow the use of classical merge tree machinery while preserving meaningful multi-field semantics.

\vspace{0.5em}\noindent\textbf{Computation of merge and contour trees.}
Foundational algorithms for contour tree computation were introduced by Carr et al.~\cite{Carr2003ContourTrees} and later refined through output-sensitive constructions~\cite{Chiang2005Monotone}. Subsequent work has focused largely on scalability, including parallel level-set topology computation~\cite{Pascucci2004parallel}, distributed merge trees ~\cite{Morozov2014Distributed}, parallel peak pruning ~\cite{Carr2021ParallelPeakPruning}, and hybrid and distributed contour-tree methods~\cite{Acharya2015ContourTree,Maadasamy2012Hybrid,Carr2022DistributedHierarchical}. Complementary approaches reduce or restructure the computation itself: triplet merge trees provide an efficient representation for constructing and manipulating merge trees~\cite{Smirnov2017TripletMergeTrees}, while ExTreeM computes merge trees on reduced extremum graphs without changing the merge tree of the original scalar field~\cite{Lukasczyk2024Extreem}. Reeb graph algorithms generalize contour-tree computation by capturing the evolution of level-set components for scalar functions on more general domains, including manifolds and non-manifold complexes ~\cite{Doraiswamy2009efficient,Doraiswamy2013ReebGraph}. A recent survey summarizes these developments with emphasis on scalable computation of topological abstractions for scalar data~\cite{will2026scalable}.

Correctness of merge trees under different interpolations is an important topic for our work. 
Topology-guided tessellation of higher-order elements~\cite{Dillard2009Quadratic} and contour tree computation for piecewise-polynomial fields~\cite{Nucha2017Polynomial} show that nonlinear interpolation can introduce topological events missed by vertex-based PL approximations. Similar issues arise for tensor-field anisotropy, whose within-cell variation is typically nonlinear~\cite{Masood2021AnisotropyHistogram}. These works emphasize the need to account for continuous variation inside cells when computing exact topological summaries~\cite{Carr2009}.

\section{Mathematical Background}
\label{sec:background}

We review the geometric and topological concepts underlying our construction
of exact trait-induced merge trees, including PL bivariate maps, inverse-image
features, Jacobi edges, and merge trees.

\vspace{0.5em}\noindent\textbf{Multifields and PL bivariate maps.}
A \emph{multifield} is a collection of $m$ scalar fields on a common $n$-dimensional spatial domain $D$. It is a multivariate map
$
    f:D \to \mathcal{A}\subseteq\mathbb{R}^m.
$
In this paper, we focus on PL bivariate maps on 3D domains, i.e., the case $n=3, m=2$. Let $M$ be a finite simplicial complex, typically a tetrahedral mesh triangulating a $3$-manifold, and let $|M|$ be its geometric realization. A
\emph{PL bivariate map} is
$
    f=(f_1,f_2):|M|\to\mathbb{R}^2,
$
where each $f_i:|M|\to\mathbb{R}$ is affine on simplices of $M$. We assume $f$ is generic: no three vertex images are collinear and no three edge images meet at a common interior point~\cite{Hristov2025Predicates}. For a vertex $v$, $f(v)$ is a \emph{vertex point}; for an edge $e=[a,b]$, $f(e)$ is the \emph{segment} joining $f(a)$ and $f(b)$.

\vspace{0.5em}\noindent\textbf{Traits, feature level sets, and fibers.}
A \emph{trait} is a subset $\tau\subseteq\mathcal{A}$ representing attribute configurations of interest. Its exact \emph{feature} is the inverse image
$
    f^{-1}(\tau)=\{x\in D:f(x)\in\tau\}.
$
To also capture nearby regions, define the distance-to-trait function
\begin{equation}
    g_\tau : |M| \to \mathbb{R},
    \qquad
    g_\tau(x)=\operatorname{dist}(f(x),\tau).
    \label{eq:trait-distance-field}
\end{equation}
where the distance is the minimum Euclidean distance from $f(x)$ to $\tau$. The level sets of $g_\tau$ are \emph{feature level sets} (FLS); the zero level set equals $f^{-1}(\tau)$, while positive levels describe approximate matches to the trait.

Specifically, for a bivariate map $f:|M|\to\mathbb{R}^2$, if $\tau=\{p\}$, then
$
    g_\tau^{-1}(0)=f^{-1}(p)=\lambda_p,
$
where $\lambda_p$ is the \emph{fiber} of $p$; its connected components are \emph{fiber components}. A simplex is \emph{active} if it intersects the fiber. For generic PL bivariate maps on tetrahedral meshes, fibers are generically one-dimensional PL curves. If $L\subseteq\mathbb{R}^2$ is a polygonal region, then $\Lambda_L=f^{-1}(L)$ is a \emph{fiber surface}. Thus, fibers and fiber surfaces are special zero-level FLS for bivariate fields.

\vspace{0.5em}\noindent\textbf{Stars, links, and Jacobi edges.}
For a simplex $\sigma\in M$, the \emph{star} and \emph{link} capture the local neighborhood $\sigma$. They are defined as
\[
    \mathrm{St}(\sigma)=\{\eta\in M:\sigma\subseteq\eta\},
    \;
    \mathrm{Lk}(\sigma)=
    \{\eta\in M:\eta\cap\sigma=\emptyset,\; \eta\cup\sigma\in M\}.
\]
In a triangulated $3$-manifold, the link of a vertex is a triangulated $2$-sphere and the link of an edge is a cycle.

Let $e=[a,b]$ be an edge. The oriented segment from $f(a)$ to $f(b)$ partitions the vertices of $\mathrm{Lk}(e)$ by their orientation in the range, inducing an \emph{upper link} $\mathrm{Lk}^+(e)$ and a \emph{lower link} $\mathrm{Lk}^-(e)$. The edge $e$ is \emph{regular} if both links are connected; otherwise it is \emph{singular}, or a \emph{Jacobi edge}. A vertex is singular if it is incident to a singular edge. The union of singular edges and singular vertices is the \emph{Jacobi set} $S_f$. Fiber topology can change only at $S_f$: definite Jacobi edges create or destroy fiber components, while indefinite Jacobi edges merge or split them.

\vspace{0.5em}\noindent\textbf{Merge trees and TIMTs.}
Let $K$ be a connected simplicial complex and let
$h:|K|\to\mathbb{R}$ be a scalar function. The \emph{merge tree} of $h$ encodes the evolution of connected components of the sublevel sets $h^{-1}((-\infty,\alpha])$ as $\alpha$ increases. Thus, it records changes in $0$-dimensional homology, but not higher-dimensional features such as loops. This definition applies equally to scalar fields on tetrahedral meshes like $M$ and to scalar functions defined on graphs (which can be viewed as 1D simplicial complexes).

For a bivariate map $f:|M|\to\mathbb{R}^2$ defined on connected domain $M$ and a trait $\tau\subseteq\mathbb{R}^2$, the \emph{trait-induced merge tree} (TIMT) is the merge tree of $g_\tau$. Equivalently, it tracks the components of
$
    M_\alpha=\{x\in |M|:g_\tau(x)\le\alpha\}
$
as $\alpha$ increases. For a point trait $\tau=\{q\}$, $g_\tau(x)=\|f(x)-q\|$, which is the main case studied in this paper.

\section{Exact Computation of TIMTs}
\label{sec:exact-timt}

A common way to compute a TIMT is to sample the distance field $g_\tau$ at the vertices of $M$ and linearly interpolate these values inside each tetrahedron, thereby obtaining a PL approximation $\widetilde g_\tau$~\cite{Jankowai2023TIMT}. This approximation is generally not exact, since the distance-to-trait function $g_\tau$ is not PL on the original tetrahedral mesh. In this section, we focus on the case of a point trait
$\tau=\{q\}$ and write
\begin{equation}
    g : |M| \to \mathbb{R},
    \qquad
    g(x)=g_\tau(x)=\|f(x)-q\|.
\end{equation}

One possible route to exact computation of TIMT for $g$ would be to subdivide each tetrahedron into monotone regions for $g$, as done in earlier work on quadratic interpolants and related nonlinear scalar fields ~\cite{Dillard2009Quadratic,Masood2021AnisotropyHistogram}. However, the present setting admits a simpler construction. First, $g$ is not an arbitrary quadratic or nonlinear interpolant: it is the Euclidean distance to a fixed point after an affine map, and its sublevel sets inside each tetrahedron has a nice structure that we describe later. Second, we seek only the merge tree, not the contour tree. Thus, we need to capture births and merges of sublevel-set components, but not split saddles or local maxima.

We therefore avoid an explicit subdivision into monotone polyhedra, eliminating the need for a more complex case-based algorithm and implementation that would increase both time and space costs. Instead, we characterize the local merge-tree structure inside each tetrahedron and then glue these local structures across shared simplices resulting in a finite weighted topological graph $G^\tau$. We show that merge tree for this graph is isomorphic to the exact TIMT of $g$ on $|M|$.

Throughout this section, unless stated otherwise, we assume distinct event values and a connected manifold tetrahedral mesh.
Also note that $f$ is affine on each simplex by the PL assumption.

\subsection{Local Merge Tree Structure}
\label{subsec:local_structure}

First, we analyze the local behavior of $g$ when restricted to a tetrahedron $T$. We show that the key local property of point-trait distance fields is that their restricted sublevel sets are convex on every simplex of the tetrahedral mesh. This gives both the triviality of the merge tree inside a tetrahedron and the completeness of the boundary events used later in the graph construction. 

\begin{lemma}[Simplex-wise convexity]
\label{lem:simplex-wise-convexity}
Let $\sigma$ be any simplex of $M$, and define
$
    \sigma_\alpha=\sigma\cap \{x:g(x)\le \alpha\}.
$
Then, for every $\alpha\ge 0$, the set $\sigma_\alpha$ is either empty or convex. In particular, it is connected whenever it is nonempty.
\end{lemma}

\begin{proof}
Since $f$ is PL, its restriction $f|_\sigma$ is affine. Hence
$
    \sigma_\alpha
    =
    \sigma\cap f^{-1}(B(q,\alpha)),
$
where $B(q,\alpha)$ is the closed Euclidean disk centered at $q$ with radius $\alpha$. The simplex $\sigma$ is convex, the disk $B(q,\alpha)$ is convex, and the preimage of a convex set under an affine map is convex. Therefore $\sigma_\alpha$ is convex as an intersection of convex sets.
\end{proof}

\begin{corollary}[Local triviality of merge tree inside tetrahedra]
\label{cor:local-triviality}
For every tetrahedron $T$ of $M$, the merge tree of $g|_T$ consists of a single monotone branch. Equivalently, no merge event can occur strictly inside a tetrahedron.
\end{corollary}

\begin{proof}
By Lemma~\ref{lem:simplex-wise-convexity}, every nonempty restricted sublevel set $T_\alpha$ is connected. Thus, as $\alpha$ increases, a single component is born at the minimum of $g|_T$ and then grows monotonically until it fills $T$. No merge event can occur within $T$.
\end{proof}

\begin{corollary}[Localization of global events]
\label{cor:boundary-events}
For a point trait, every global change in sublevel-set connectivity can occur only through shared simplex boundaries, namely through vertices, edges, or faces shared by tetrahedra.
\end{corollary}

\begin{proof}
By Corollary~\ref{cor:local-triviality}, the interior of each tetrahedron contributes only one growing connected component. Hence a global connectivity change can only occur when components in adjacent tetrahedra attach through their common intersection, which is a shared face, edge, or vertex.
\end{proof}

\begin{figure}[h]
    \centering

    \begin{subfigure}[t]{\linewidth}
        \centering
        \includegraphics[width=\linewidth]{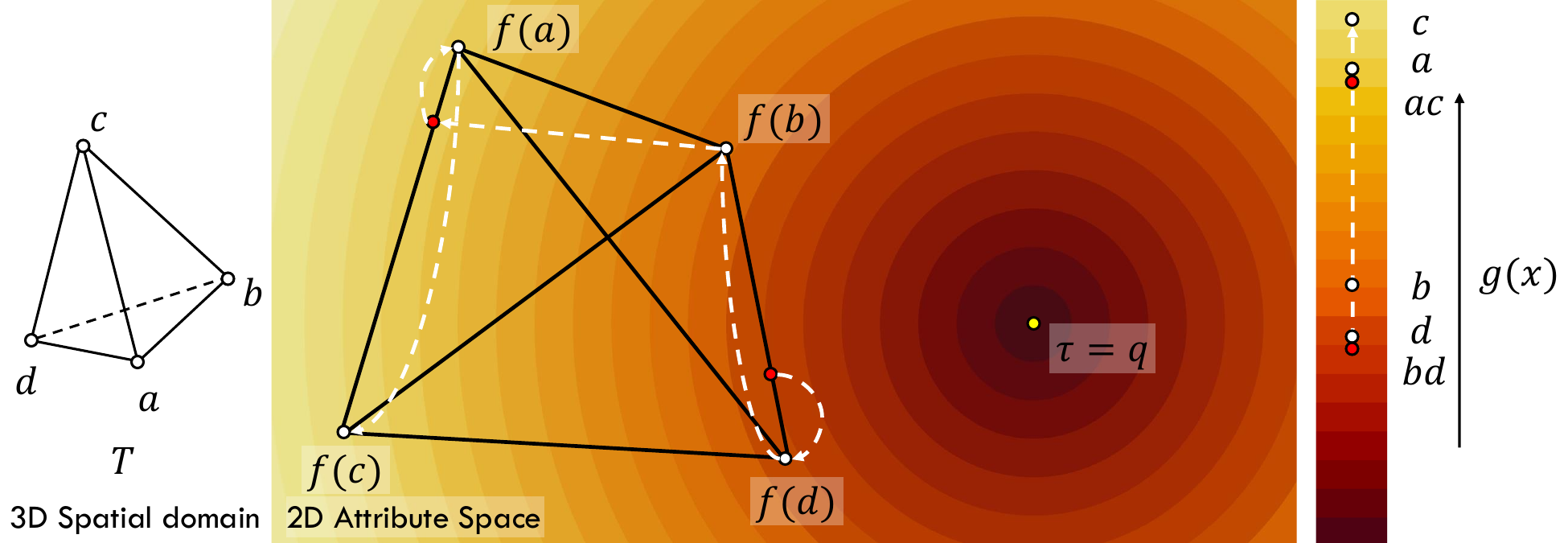}
        \label{fig:local-merge-tree-a}
        (a)\quad
    \end{subfigure}

    \vspace{0.35em}

    \begin{subfigure}[t]{\linewidth}
        \centering
        \includegraphics[width=\linewidth]{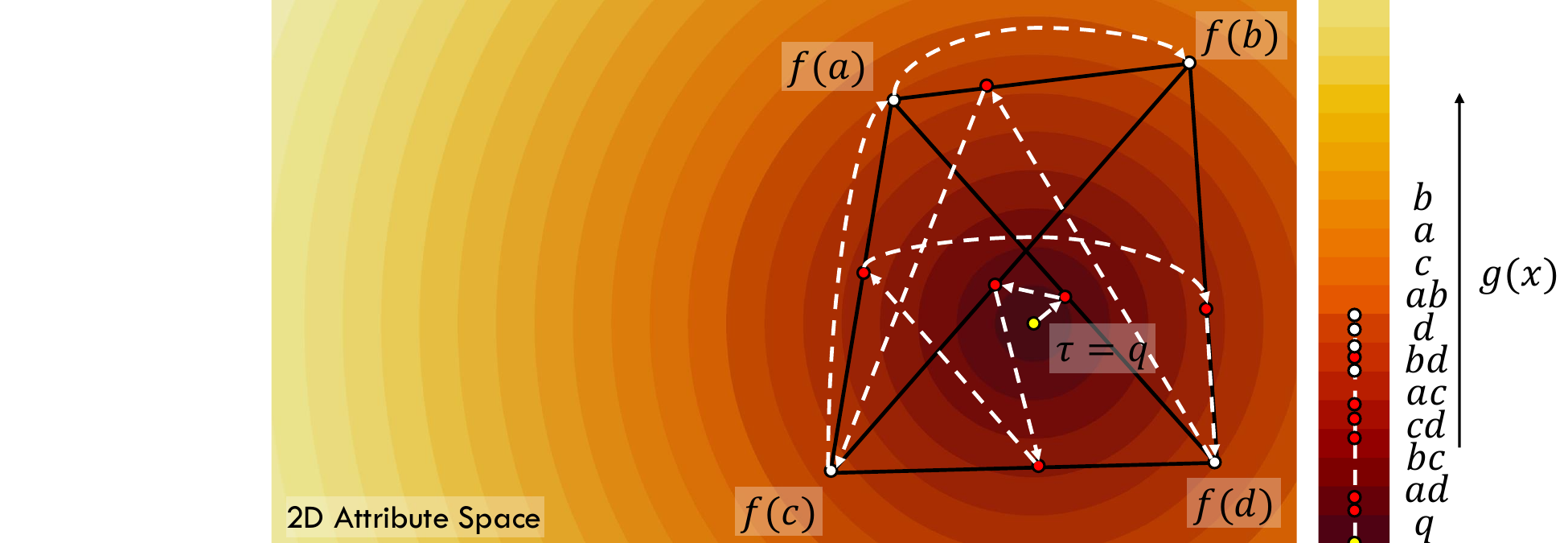}
        \label{fig:local-merge-tree-b}
        (b)\quad
    \end{subfigure}

    \vspace{0.35em}

    \begin{subfigure}[t]{\linewidth}
        \centering
        \includegraphics[width=\linewidth]{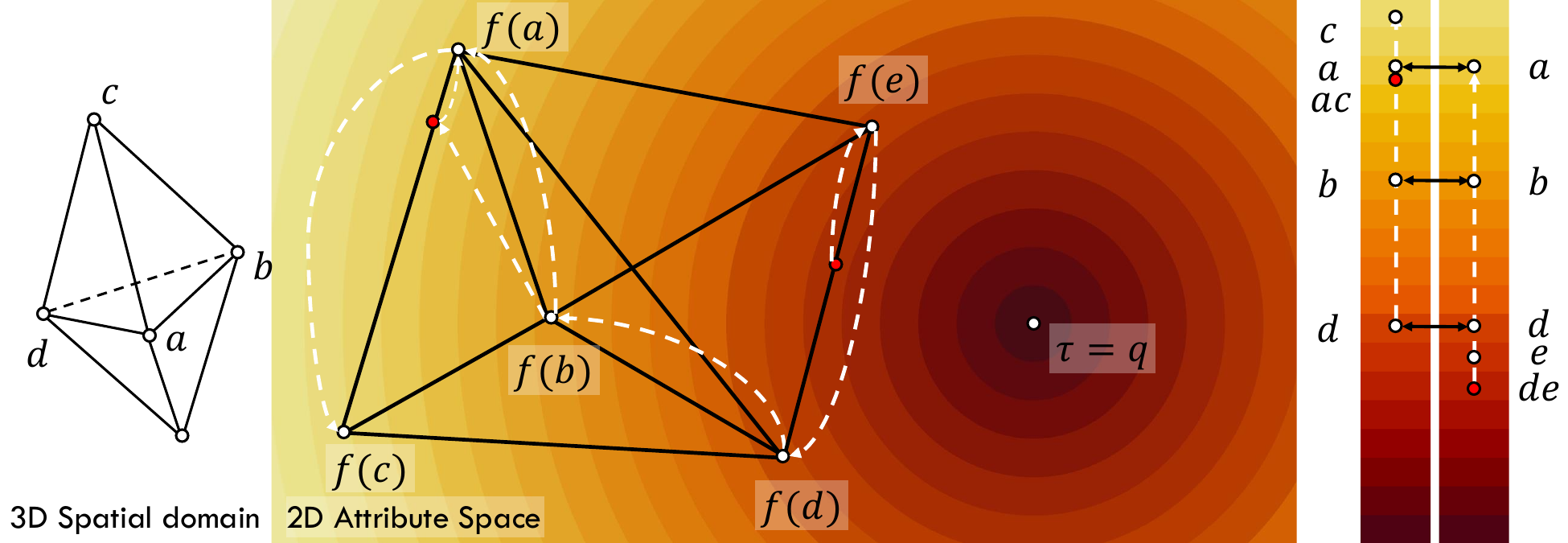}
        \label{fig:local-merge-tree-c}
        (c)\quad
    \end{subfigure}

    \caption{Local construction of the trait-induced graph. (a) and (b) illustrate the trivial local merge-tree structure inside individual tetrahedra: the nodes are ordered by increasing value and connected into a monotone chain, shown by dashed arrows. In case (b) the trait point lies inside $f(T)$; the zero-fiber node (yellow) enters first, and the tetrahedron contains one positive-valued edge-minimum event for each of its six edges. 
    (c) shows two tetrahedra sharing a face. Their local chains in the trait-induced graph are glued by identifying nodes on the common boundary.}
    \label{fig:local-trait-graph}
\end{figure}

The same simplex-wise convexity also applies to shared boundaries. Let $\sigma$ be a shared simplex with $\textnormal{dim}(\sigma) \leq 2$. Since $\sigma_\alpha$ is empty or connected, its first nonempty value is the only possible attachment event through $\sigma$. If $\sigma$ is a vertex, this event is the vertex itself. If $\sigma$ is an edge, the first point is either an endpoint or a positive-valued interior minimum of $g|_\sigma$. If $\sigma$ is a face, the first point is the closest point to $q$ in the range triangle $f(\sigma)$; if this closest point lies in the interior, the event has value zero and belongs to the fiber $f^{-1}(q)$, while if it lies on the boundary, it reduces to an edge or vertex event.
Thus, every boundary attachment that can change sublevel-set connectivity is represented by one of three types of nodes: an input mesh vertex, a positive-valued interior edge minimum, or a component of the fiber $f^{-1}(q)$. See \cref{fig:local-trait-graph} for illustrations of local merge tree and attachment across shared faces.

\subsection{The Trait-induced Graph}
\label{subsec:full-trait-graph}

Our analysis above shows that, for a point trait, each tetrahedron contributes only one connected sublevel component, while global connectivity changes at shared simplices. 
Using this, we construct a finite weighted graph $G^\tau$ called \emph{trait-induced graph} whose merge tree is exact for the point-trait distance field $g$.
The nodes of $G^\tau$ are:

\begin{enumerate}[noitemsep]
    \item input mesh vertices;
    \item positive-valued interior minima of $g$ on mesh edges; and,
    \item a node each for connected components of the fiber $f^{-1}(q)$.
\end{enumerate}
Each node is assigned its corresponding value of $g$, with zero-fiber nodes assigned weight $0$.

The construction is summarized in \cref{alg:trait-graph}, and a detailed version is provided in the appendix. For each tetrahedron $T$, we collect all nodes incident to $T$, sort them by nondecreasing weight, and connect consecutive nodes by graph edges. Nodes supported on shared mesh entities are identified globally, so each mesh vertex, edge-minimum node, and zero-fiber component has a unique representative across all incident tetrahedra. The running time complexity of \cref{alg:trait-graph} is linear in the number of simplices of the input mesh $M$. Each simplex is processed once, and all associated operations are performed in constant time. The only non-trivial step is sorting the nodes within a tetrahedron, but this cost is also bounded because at most $11$ nodes can be incident on a tetrahedron. Consequently, the overall complexity of the algorithm is $O(|M|)$.

$G^\tau$ is a weighted graph whose node weights are extended monotonically along each edge. The graph sublevel set $G^\tau_\alpha$ is the subgraph consisting of points with interpolated value at most $\alpha$.

\begin{algorithm}[t]
\caption{Construction of the exact trait-induced graph $G^\tau$}
\label{alg:trait-graph}
\begin{algorithmic}[1]
\Require Tetrahedral mesh $M$, PL bivariate field
$f:M\to\mathbb{R}^2$, point trait $q\in\mathbb{R}^2$
\Ensure Weighted graph $G^\tau=(V,E)$ whose node weights are induced by
$g(x)=\|f(x)-q\|$

\State $V \gets \emptyset$, $E \gets \emptyset$

\ForAll{vertices $v$ of $M$}
    \State add node $n_v$ to $V$ with weight $g(v)$
\EndFor

\ForAll{edges $e$ of $M$ \textbf{in parallel}}
    \If{$g|_e$ has a positive-valued interior minimum at $p_e$}
        \State add node $n_e$ to $V$ with weight $g(p_e)$
        \State record $\textsc{EdgeNode}[e] \gets n_e$
    \EndIf
\EndFor

\State compute the connected components of the zero fiber $f^{-1}(q)$
\ForAll{zero-fiber components $C$}
    \State add node $n_C$ to $V$ with weight $0$
    \State record the tetrahedra intersected by $C$ as incident to $n_C$
\EndFor

\ForAll{tetrahedra $T$ of $M$ \textbf{in parallel}}
    \State let $V_T$ be the set of nodes incident to $T$
    \State sort $V_T$ by nondecreasing weight
    \For{$i=1$ to $|V_T|-1$}
        \State add edge $(V_T[i],V_T[i+1])$ to $E$
    \EndFor
\EndFor

\State \Return $G^\tau=(V,E)$
\end{algorithmic}
\end{algorithm}

\begin{lemma}[Exactness]
\label{lem:graph-exactness}
For every regular value $\alpha$, the connected components of $G^\tau_\alpha$ are in one-to-one correspondence with those of
$
    M_\alpha=\{x\in |M|:g(x)\le \alpha\}.
$
Consequently, the merge tree of $G^\tau$ is isomorphic to the exact TIMT of $g$ on $|M|$.
\end{lemma}

\begin{proof}
Graph edges are added only between consecutive nodes in sorted order incident to a common tetrahedron. By Lemma~\ref{lem:simplex-wise-convexity}, the corresponding restricted sublevel sets are connected between consecutive event values. Thus, every connection in $G^\tau_\alpha$ represents a valid sublevel set nesting relationship in $M_\alpha$.

Conversely, any path in $M_\alpha$ crosses tetrahedra through shared faces, edges, or vertices. By Corollary~\ref{cor:boundary-events} and the boundary-event classification following Lemma~\ref{lem:simplex-wise-convexity}, each such attachment is represented in $G^\tau$ by a mesh vertex, an edge-minimum node, or a zero-fiber node. Every
connection in $M_\alpha$ is realized in $G^\tau_\alpha$ so
$G^\tau_\alpha$ and $M_\alpha$ have the same connected components for every regular value $\alpha$ and their merge trees are isomorphic.
\end{proof}

\subsection{Robust Computation and Ordering of Local Nodes}
\label{subsec:robust-edge-minima}

The construction of $G^\tau$ requires two key operations: detecting the positive-valued edge minima of $g$ and ordering the incident graph nodes within each tetrahedron. Here we describe how to perform them in a robust way.

Let $e=[a,b]$ be an edge of $M$. Since $f$ is affine on $e$, the image $f(e)$ is the segment joining $f(a)$ and $f(b)$ in $\mathbb{R}^2$. Thus, detecting an interior minimum of $g(x)=\|f(x)-q\|$ on $e$ is equivalent to finding the closest point from $q$ to the segment $f(e)$. Let
\[
    u=f(b)-f(a), \qquad v=q-f(a).
\]
The orthogonal projection of $q$ onto the supporting line of $f(e)$ lies in the interior of the segment if and only if
\begin{equation}
    0 < \langle v,u\rangle < \langle u,u\rangle .
\end{equation}
This test avoids computing the projection parameter explicitly. When the condition holds, the edge-minimum point is given by
\[
    p_e = (1-t)a + tb,
    \qquad
    t=\frac{\langle v,u\rangle}{\langle u,u\rangle},
\]
and its squared distance value is
\begin{equation}
    g(p_e)^2
    =
    \|v\|^2
    -
    \frac{\langle v,u\rangle^2}{\langle u,u\rangle}.
\end{equation}
We add an edge-minimum node only when this value is positive; the zero case is represented instead by the corresponding component of the fiber $f^{-1}(q)$.

After all local nodes incident to a tetrahedron have been identified, they are sorted by their scalar value. The ordering is constrained by two structural facts. First, if a zero-fiber node is incident to the tetrahedron, it has value zero and must appear first. Second, if an edge-minimum node is present on an edge $[a,b]$, then its value is strictly smaller than the values at both endpoints:
\[
    g(p_e) < g(a), \qquad g(p_e) < g(b).
\]
Thus, the edge-minimum node must appear before the two corresponding vertex nodes in the local order.

These constraints are sufficient for the construction of $G^\tau$. In particular, we do not require an additional simulation of simplicity scheme at this stage. The remaining ties between nodes of equal value that do not violate the structural constraints above do not affect the construction of the weighted graph; they can be handled by the downstream merge-tree algorithm or library. Our output is therefore a weighted graph with the correct local ordering implied by zero-fiber nodes and edge minima, while any further tie-breaking is left to the merge-tree computation.

\subsection{TIMTs and Jacobi Set}
\label{subsec:timt-jacobi}

We now relate TIMT critical events to the Jacobi set of the bivariate map $f$. The key observation is that a positive-valued edge minimum can be critical for the TIMT only if crossing the corresponding edge changes the local connectivity of fibers, which is precisely the role of Jacobi edges.

\begin{lemma}[TIMT-critical edge minima lie on Jacobi edges]
\label{lem:timt-critical-edge-jacobi}
Let $e$ be a mesh edge and let $p\in e$ be a positive-valued interior minimum of $g|_e$. If $p$ is a critical event of the TIMT of $g(x)=\|f(x)-q\|$, then $e$ is a Jacobi edge of $f$.
\end{lemma}

\begin{proof}
Let $y=f(p)$ and $r=g(p)=\|y-q\|>0$.
Since $p$ is an interior minimum of $g|_e$, the point $y$ is the closest point from $q$ to the segment $f(e)$ in the range. Thus, as $\alpha$ crosses $r$, the disk $B(q,\alpha)$ first touches the segment $f(e)$ at the point $y$.

Consider the local evolution of the sublevel sets $M_\alpha=f^{-1}(B(q,\alpha))$
in the star of $e$. If $p$ is a TIMT minimum, then a new local component of $M_\alpha$ is born as $\alpha$ passes through $r$. Equivalently, a fiber component is created at the edge, which is the behavior of a definite Jacobi edge. If $p$ is a TIMT merge saddle, then two or more local components of $M_\alpha$ become connected when $\alpha$ crosses $r$. Equivalently, fiber components merge at the edge, which is the behavior of an indefinite Jacobi edge.

In both cases, crossing $f(e)$ changes the local connectivity of fibers of the bivariate map. Hence, the edge fails the regular-edge criterion and is a Jacobi edge.
\end{proof}

Lemma~\ref{lem:timt-critical-edge-jacobi} identifies where nonzero edge-based TIMT events can occur. For point traits, nonzero minima or merge saddles of the TIMT may occur at input mesh vertices or at positive-valued interior minima on Jacobi edges. Zero-valued events are represented separately by the connected components of the zero fiber $f^{-1}(q)$.

\subsection{Error Between Exact and Approximate TIMTs}
\label{subsec:exact-approx-error}

In this section, we establish bounds on the difference between the exact TIMT and the merge tree obtained from a vertex-sampled approximation of the trait-induced distance field. We do so by bounding the difference between the exact field $g(x)$ and its linear interpolation $\widetilde{g}(x)$.

Let $T=[v_0,v_1,v_2,v_3]$ be a tetrahedron, let $y_i = f(v_i)$, and $z_i = y_i - q$.
For $x \in T$ with barycentric coordinates $\lambda_i$, define $m = f(x)-q = \sum_i \lambda_i z_i$ as its projection in range space.
Then
\[
    g(x) = \|m\|= \|\sum_i \lambda_i z_i\|,
    \qquad
    \widetilde{g}(x) = \sum_i \lambda_i \|z_i\|.
\]

First, by convexity of the norm, $g(x) \le \widetilde{g}(x)$, so the error is non-negative. In particular, the sublevel sets satisfy the following
$
    \{x : \widetilde{g}(x) \le \alpha\}
    \subseteq
    \{x : g(x) \le \alpha\}.
$

Using the triangle inequality, for each $i$, $\|z_i\| - \|m\| \le  \|z_i - m\|$,
which implies
\[
\begin{aligned}
    \widetilde{g}(x) - g(x)
    &= \sum_i \lambda_i \|z_i\| - \|m\| \\
    &\le \sum_i \lambda_i \|z_i - m\| \\
    &\le \max_i \|z_i - m\| \le \max_{i,j} \|z_i - z_j\| = \max_{i,j} \|y_i - y_j\|.
\end{aligned}
\]

Or, for the whole mesh $M$, the error is bounded by the maximum length of edges in range space:
\begin{equation}
    \varepsilon = \|g - \widetilde{g}\|_\infty
    \le \max_{e=(v_i,v_j) \in E(M)} \|f(v_i)-f(v_j)\|.
\end{equation}

\begin{figure}[!htb]
    \centering
    \includegraphics[width=.9\linewidth]{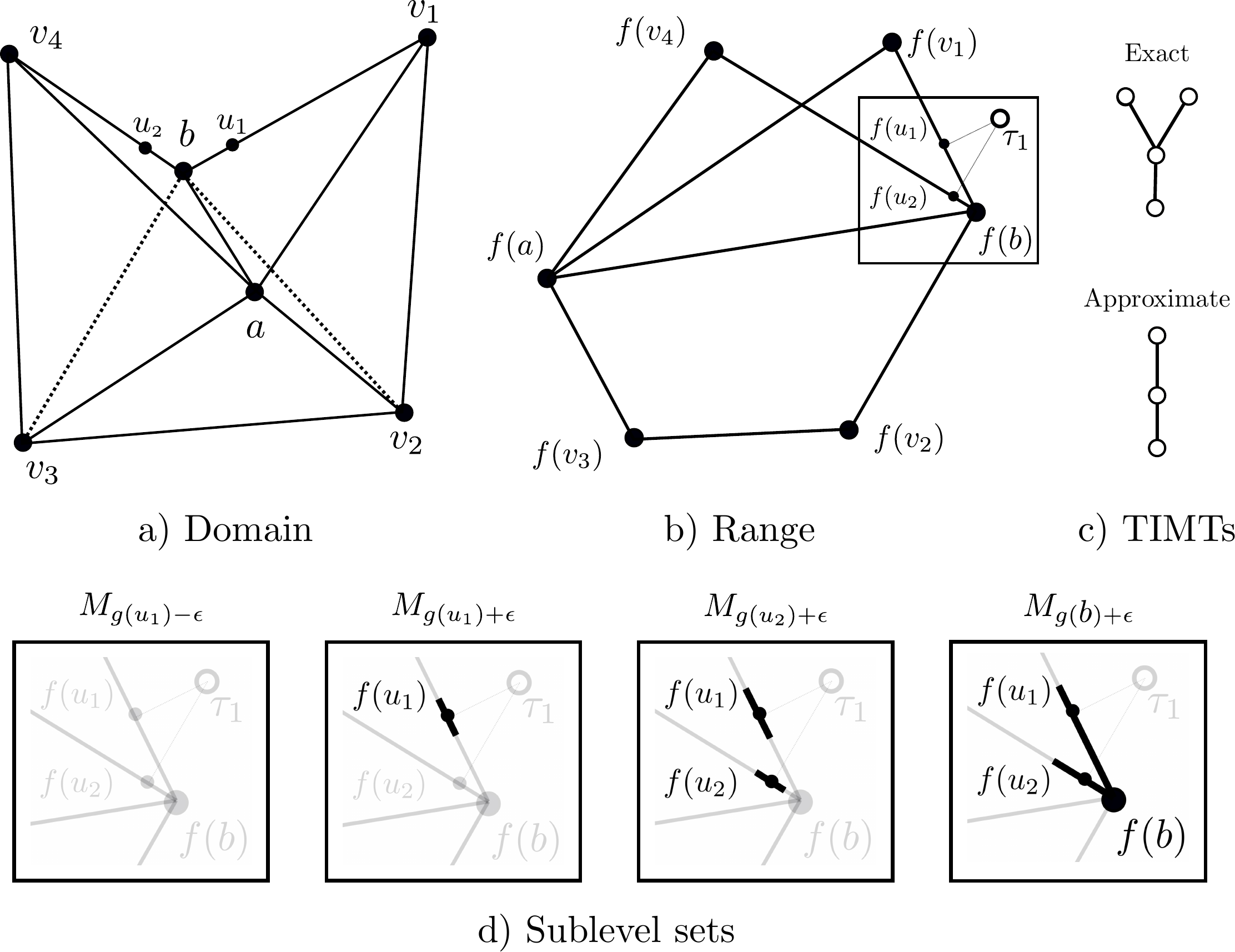}
    \caption{Differences between exact and approximate TIMTs for a point trait $\tau$ for small tetrahedral mesh.
    The domain a) has three tetrahedra around an edge $ab$.
    In the range (b) we have mapped the Jacobi edges, where edge minima lie.
    The closest points to $\tau$ are $f(u_1)$ and $f(u_2)$, the images of the points $u_1$ and $u_2$ along the edges $bv_1$ and $bv_4$ in the domain.
    Taking the images of four sublevel sets of $g$ in the range demonstrates that two components are born that merge, leading to a TIMT with a Y shape.
    The vertex with the smallest distance value is $b$ so all sublevel sets of the approximation of $g$ have a single component, leading to merge tree with one edge. 
    }
    \label{fig:toy-example}
\end{figure}

The sublevel sets, therefore, satisfy the nesting relation
\begin{equation}
    \{x : \widetilde{g}(x) \le \alpha\}
    \subseteq
    \{x : g(x) \le \alpha\}
    \subseteq
    \{x : \widetilde{g}(x) \le \alpha+\varepsilon\}.
\end{equation}
The exact and approximate filtrations are $\varepsilon$-interleaved, and the corresponding TIMTs differ by at most $\varepsilon$ in interleaving merge-tree distance~\cite{Beketayev2014Interleaving}.
For a simple illustration of differences between exact and approximate TIMTs, see \cref{fig:toy-example}.

\section{Outlook and Extensions}
\label{sec:outlook-extensions}

While the exact construction in \cref{sec:exact-timt} was developed for point traits, the underlying principle is more general. If restricted sublevel sets remain connected inside each simplex, then the merge tree can be recovered by representing the boundary events through which local components attach. The cases discussed below outline possible extensions of this principle. They should be viewed as directions for future work; each requires a complete correctness proof, robust implementation, and empirical testing.

\subsection{Beyond 3D Domains}

The point-trait construction extends naturally to bivariate PL maps $f:|M|\to\mathbb{R}^2$ whose domain has dimension $n\ge 2$. The key convexity argument is independent of the dimension of the domain: for any simplex $\sigma\in M$,
$
    \sigma_\alpha
    =
    \sigma\cap f^{-1}(B(q,\alpha))
$
is the intersection of a simplex with the affine preimage of a disk, and is therefore empty or convex. Hence, the restricted sublevel set inside each simplex is connected whenever nonempty.

\begin{figure}[!ht]
    \centering

    \begin{subfigure}[t]{\linewidth}
        \centering \includegraphics[width=0.99\linewidth]{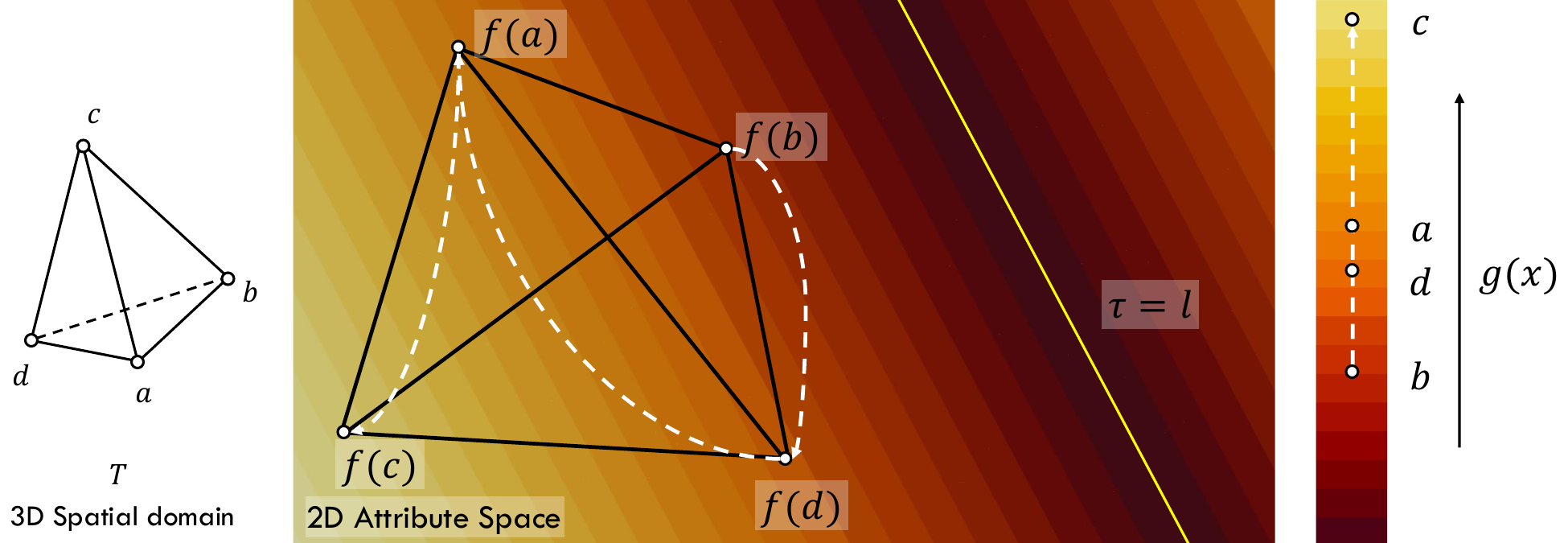}
        \label{fig:extensions-line}\\
        (a)\quad
    \end{subfigure}

    \vspace{0.35em}

    \begin{subfigure}[t]{\linewidth}
        \centering \includegraphics[width=0.99\linewidth]{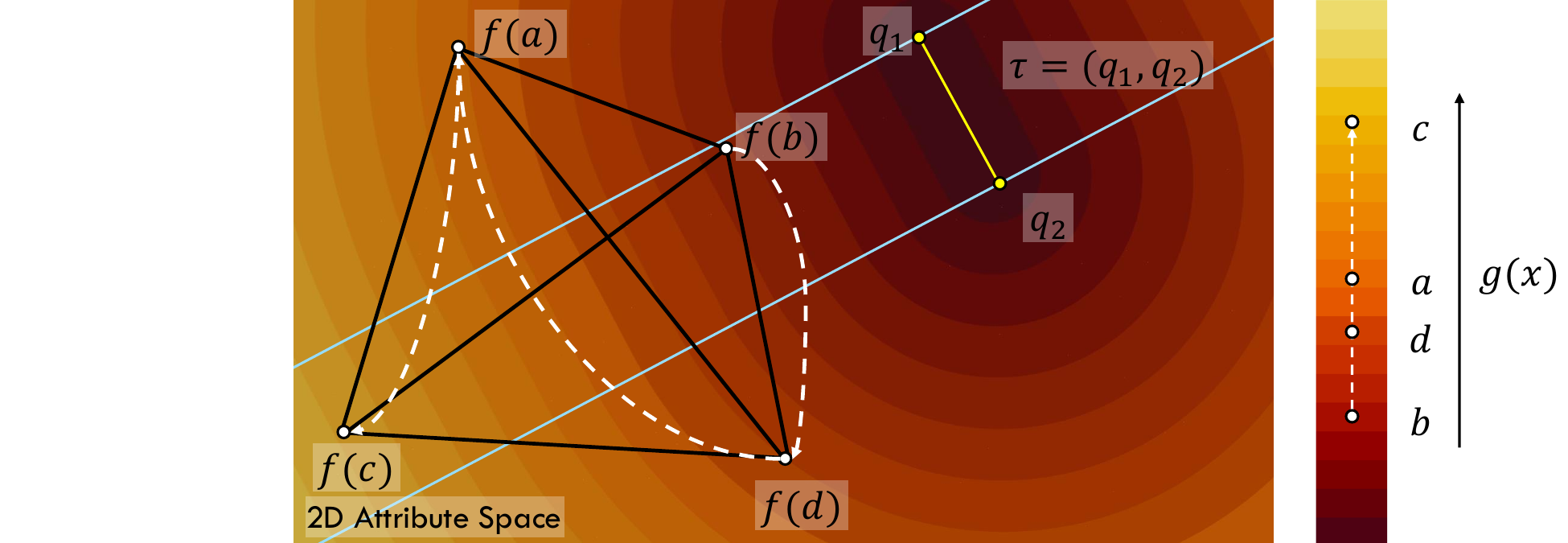}
        \label{fig:extensions-line-segment}\\
        (b)\quad
    \end{subfigure}

    \vspace{0.35em}

    \begin{subfigure}[t]{\linewidth}
        \centering \includegraphics[width=0.99\linewidth]{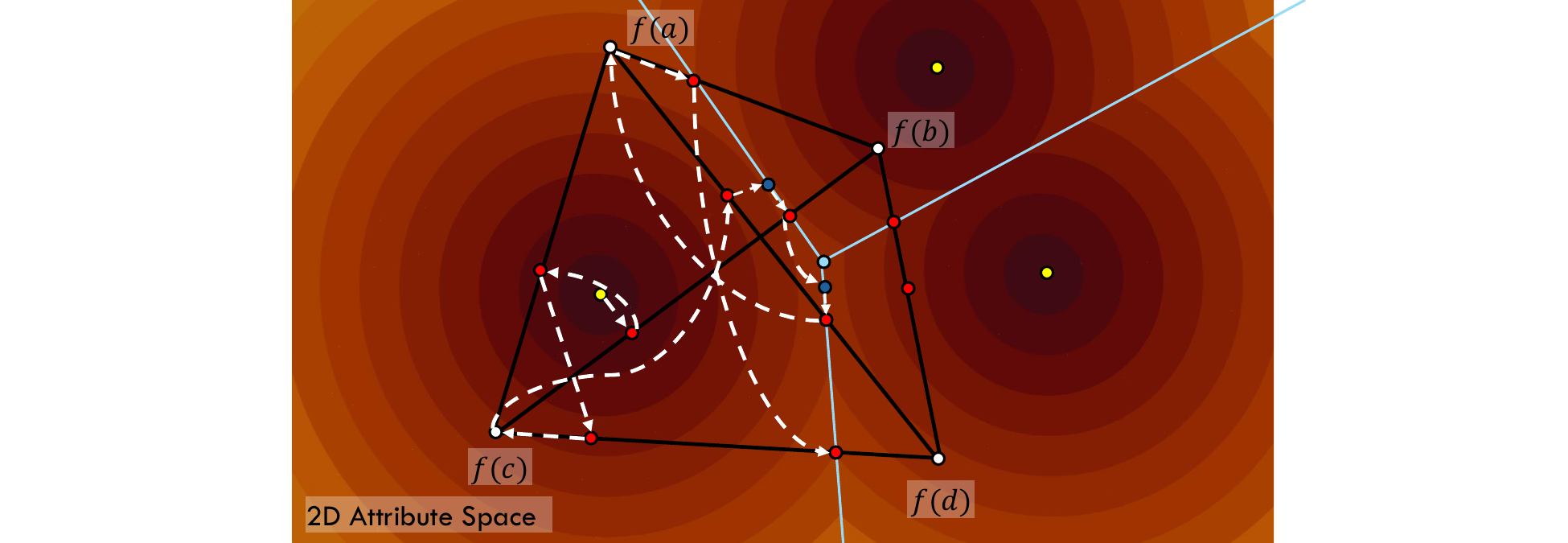}
        \label{fig:extensions-point-set}\\
        (c)\quad
    \end{subfigure}

    \caption{Attribute space decompositions for extending the exact TIMT
    construction beyond point traits. 
    (a) For a line trait, the sublevel sets in the range are strips. Their intersection with $f(T)$ induces a PL distance field on the tetrahedron, so the local construction reduces to ordering the vertices by their distance value. 
    (b) For a line-segment trait, the attribute space decomposes into three convex regions: two endpoint regions, where the closest point is one of the segment endpoints, and a middle region, where the closest point lies in the segment interior. Intersecting these regions with $f(T)$ yields convex pieces; the endpoint pieces reduce to the point-trait construction, while the middle piece follows the line-trait construction. 
    (c) For a finite point-set trait, the Voronoi diagram partitions $f(T)$ into convex regions according to the closest trait point. The construction reduces locally to the point-trait case inside each Voronoi region, but additional events arise from intersections of Voronoi edges with projected mesh edges and from minima along Voronoi edges, which can be characterized through intersections between Delaunay and Voronoi edges.}
    \label{fig:extensions}
\end{figure}

For positive-valued events, the situation is almost the same as in the tetrahedral case. If $q\notin f(\sigma)$, then the closest point in the convex polygon $f(\sigma)\subset\mathbb{R}^2$ is on its boundary. The corresponding event is represented by a projected vertex or edge of $\sigma$. Thus, even in higher-dimensional domains, positive critical candidates for the TIMT can still be captured by subdividing only edges.

The zero level set requires additional care. Generically, for a bivariate map from an $n$-dimensional domain, the inverse image $f^{-1}(q)$ has dimension $n-2$. Thus, in two-dimensional domains it consists of isolated points, in three-dimensional domains it consists of curves, and in higher dimensions it becomes higher-dimensional. The graph construction can still be applied provided the connected components of the zero level set are identified correctly and represented as zero-valued nodes.

\subsection{Beyond Bivariate Maps}

The same convexity principle also applies to PL maps $f:|M|\to\mathbb{R}^m$ with higher-dimensional range. For a point trait $q\in\mathbb{R}^m$, the sublevel set in range is the ball $B(q,\alpha)$, and its affine preimage inside any simplex is convex. Hence local sublevel sets are again connected, and the local merge tree remains trivial.

The main difference is the boundary-event structure. If $q\notin f(\sigma)$, the closest point to $q$ in the convex polytope $f(\sigma)\subset \mathbb{R}^m$ may lie on a face of dimension up to $m-1$. Therefore, in addition to edge minima, we identify minimum points on higher-dimensional simplices up to dimension $m-1$ and ensure that their attachments are represented consistently in the graph. The zero level set must be handled according to its expected dimension: generically, $f^{-1}(q)$ has codimension $m$ in an $n$-dimensional domain, and therefore dimension $n-m$. Thus, while the local convexity proof extends directly, the event detection and boundary bookkeeping become more involved.

\subsection{Beyond Point Traits}

Other trait classes can be approached by decomposing the range into regions where the distance function reduces to a simpler case. We briefly discuss extensions to three class below, and refer readers to \cref{fig:extensions} for illustrations.

\vspace{0.5em}\noindent\textbf{Line traits.}
For an affine line trait in $\mathbb{R}^2$, sublevel sets in the range are strips bounded by parallel lines. Their intersections with convex $f(\sigma)$ remain convex, so the local connectedness argument carries over. Equivalently, the distance to a line is the absolute value of an affine function on each simplex. If the zero set intersects a simplex, the corresponding zero-level fiber surface must be represented. Otherwise, the distance is affine up to sign on that simplex, and no positive-valued interior edge minimum is needed.

\vspace{0.5em}\noindent\textbf{Line-segment traits.}
For a line segment with endpoints $q_1$ and $q_2$, the range decomposes into three regions: points closest to $q_1$, points closest to $q_2$, and points whose closest point lies in the interior of the segment. The endpoint regions reduce to point-trait cases, while the middle region reduces to the line-trait case. A graph construction must therefore include additional interface events where the closest-point map changes from endpoint-like to line-like behavior.

\vspace{0.5em}\noindent\textbf{Point-set traits.}
For a finite point set $Q\subset\mathbb{R}^2$, the distance field is $g_\tau(x)=\min_{q\in Q}\|f(x)-q\|.$
The range is naturally subdivided by the Voronoi diagram of $Q$. Within a Voronoi cell, the distance reduces to the distance from a single point. Consequently, the point-trait construction applies locally on each intersection of a Voronoi cell with $f(\sigma)$. Additional critical events arise only on Voronoi edges and vertices, where multiple trait points are equidistant. Incorporating these events enables the exact computation of TIMTs for point-set traits.

Overall, these extensions suggest that the exact TIMT framework is not limited to point traits or tetrahedral domains. The central requirement is to identify a finite set of local and boundary events that preserves the connectivity of sublevel sets under gluing. Formalizing this requirement for broader classes of domains, ranges, and traits remains an important direction for future work.

\section{Evaluation}
\label{sec:evaluation}

In this section, we discuss our implementation of  \cref{alg:trait-graph} and perform an evaluation of exact and approximate~\cite{Jankowai2023TIMT,Lei2025TraitDesign} computation of TIMTs.

\subsection{Implementation}

We implemented the computation of the trait-induced graph as a proof of concept application for point traits in C++  using VTK 9.3 and CGAL 6.0.
We used CGAL for all geometric computation with an exact predicates and exact constructions (EPEC) kernel~\cite{goodmanHandbookDiscreteComputational2018} to ensure the correctness of our implementation by avoiding floating-point numerical errors.
The code is publicly available at \href{https://github.com/tbmasood/Exact_TIMT}{\texttt{https://github.com/tbmasood/Exact\_TIMT}}.
For our evaluation, we used Paraview and TTK~\cite{masood2021overview} to compute merge trees and perform simplification.
We used persim~\cite{persim2025} and gudhi~\cite{gudhi2014} to compute the Wasserstein distances between persistence diagrams.

\subsection{Datasets}

\begin{figure*}
    \centering
    \includegraphics[width=0.8\linewidth]{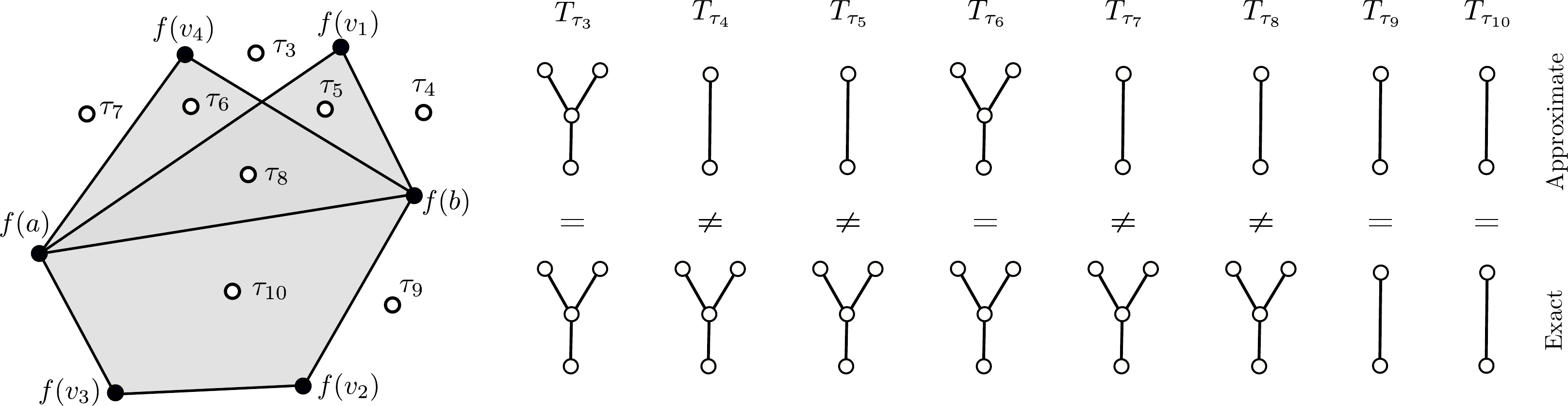}
    \caption{Differences between exact and approximate TIMTs for various traits for the three tetrahedra toy data set.}
    \label{fig:toy-evaluation}
\end{figure*}

The first dataset that we evaluated is the three tetrahedra toy example that has been used in numerous multivariate topology papers~\cite{Hristov2025ATAlgorithm, tiernyJacobiFiberSurfaces2017} and in \cref{fig:toy-example}.
This dataset clearly illustrates exactly what the differences are in the exact and approximate TIMTs.

\begin{figure*}[p]
     \centering
     \begin{subfigure}[b]{0.30\textwidth}
         \includegraphics[width=\columnwidth]{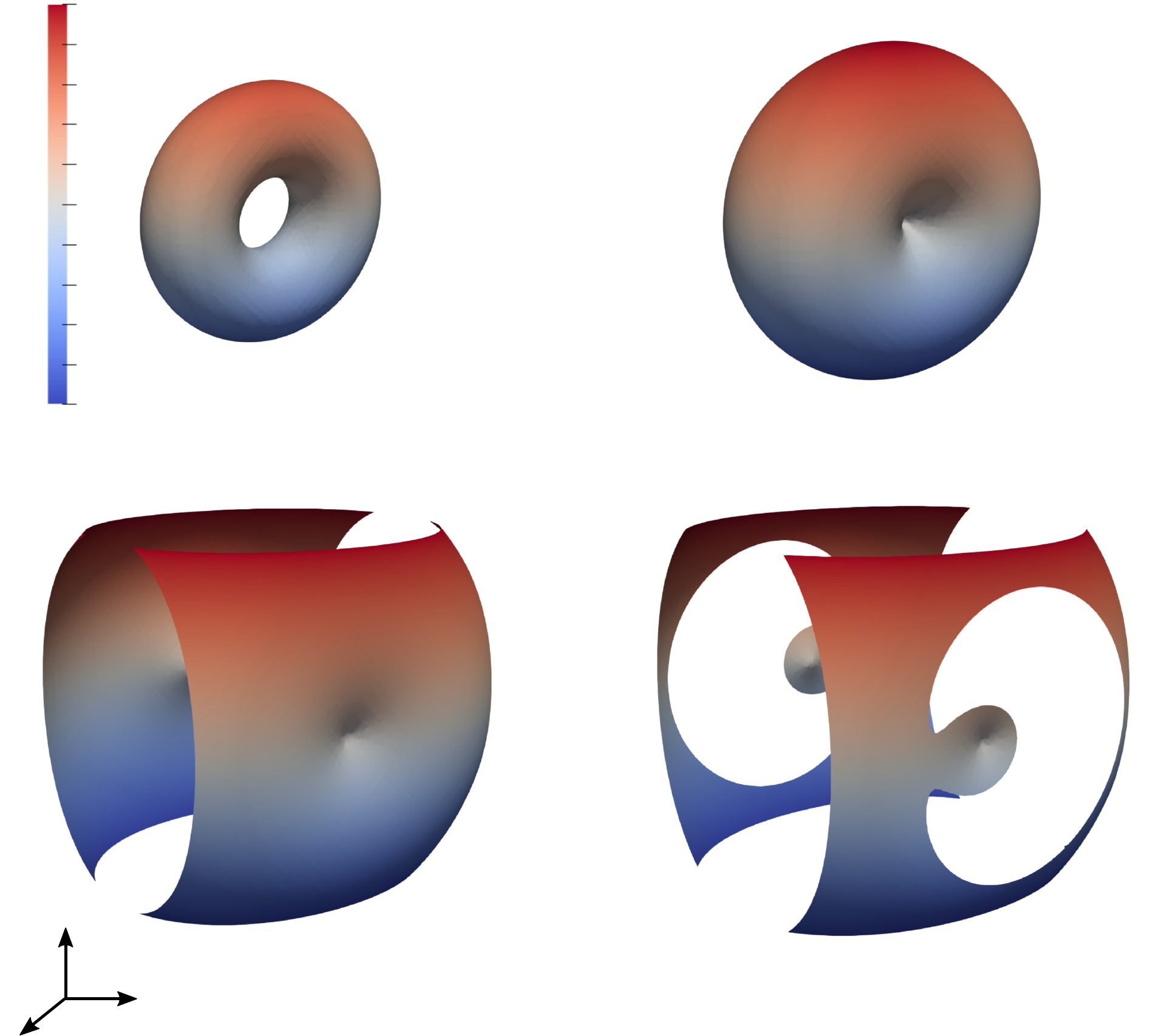}
         \caption{Isosurfaces of $f_1$ colored by $f_2$.}
     \end{subfigure}
     \begin{subfigure}[b]{0.40\textwidth}
         \includegraphics[width=\columnwidth]{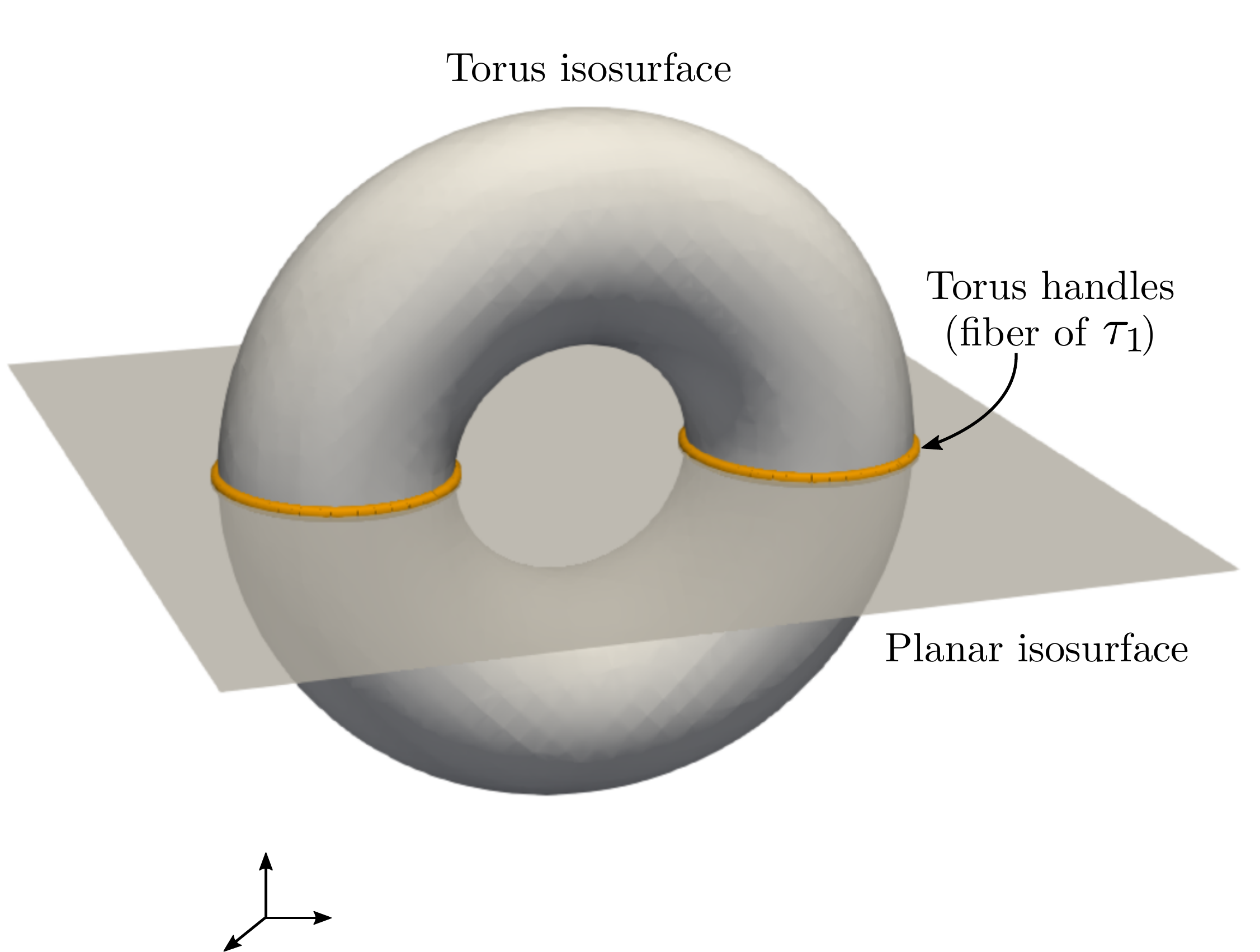}
         \caption{Isosurfaces and fiber.}
     \end{subfigure}
     \begin{subfigure}[b]{0.28\textwidth}
         \includegraphics[width=\columnwidth]{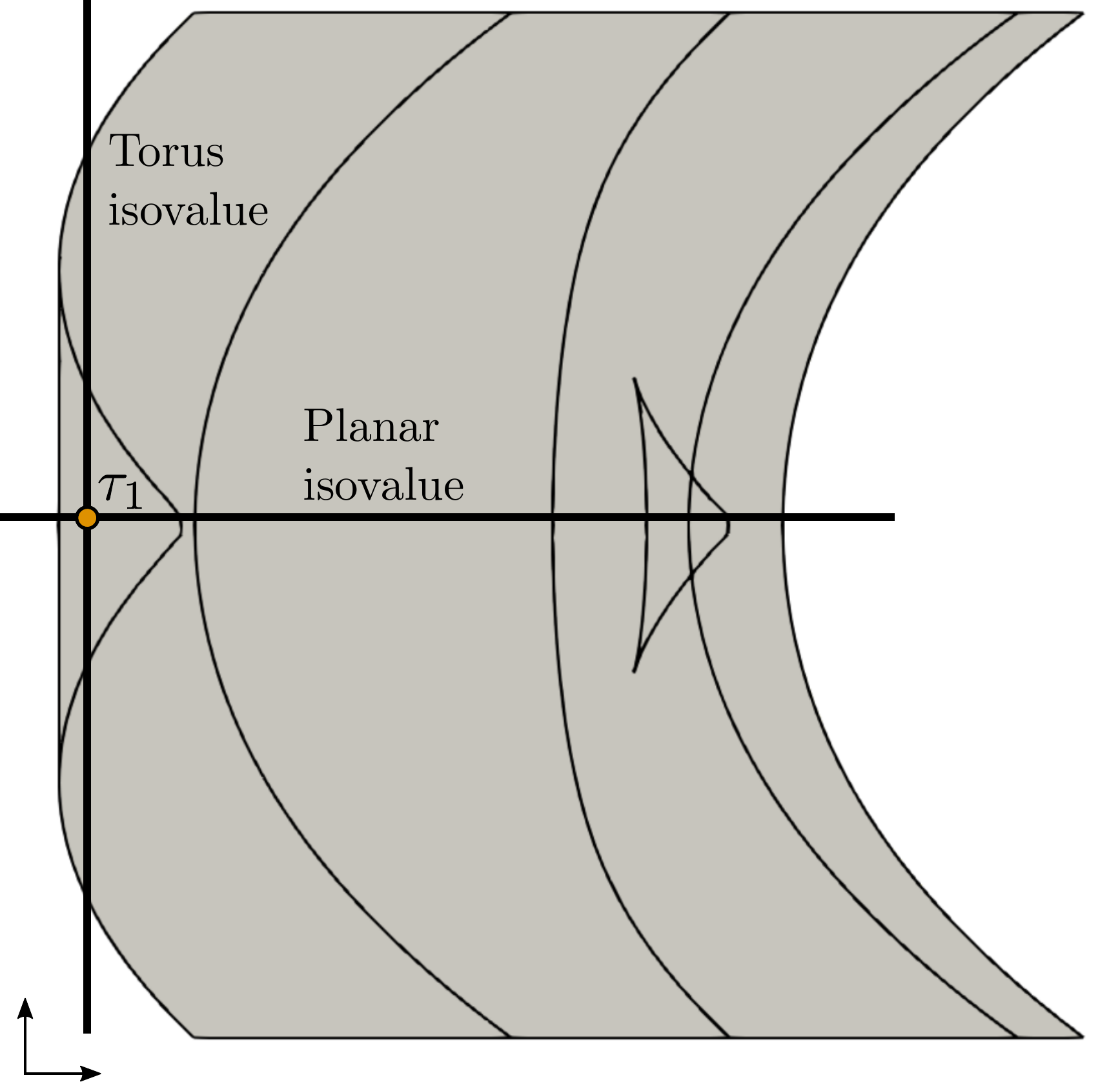}
         \caption{Range space, singular segments and trait.}
     \end{subfigure}
     \caption{
        Bivariate torus dataset.
        In (a), we show isosurfaces of the torus function $f_1$ colored by the height function $f_2$.
        In (b) we show two selected isosurfaces of $f_1$ and $f_2$, whose intersection is a fiber (colored orange), which captures the two side handles of the torus.
        In the range space (c) we render the image of the domain and the singular (Jacobi) edges.
        Here the isovalues are vertical and horizontal lines, whose intersection is the trait point $\tau_1$.
     }
     \label{fig:torus-dataset}
\end{figure*}

The second data set is the bivariate torus data set~\cite{Hristov2025ATAlgorithm} shown in \cref{fig:torus-dataset}.
The first scalar function $f_1$ in this dataset is an implicit torus function, defined so that the isosurfaces are those of an expanding torus.
The second function $f_2$ is a height function whose isosurfaces are horizontal planes.
One of the key features in this data set is the two side handles of the torus, shown in \cref{fig:torus-dataset} (b) and (c).

\begin{figure}[h]
     \centering
     \begin{subfigure}[b]{0.24\textwidth}
         \includegraphics[width=\columnwidth]{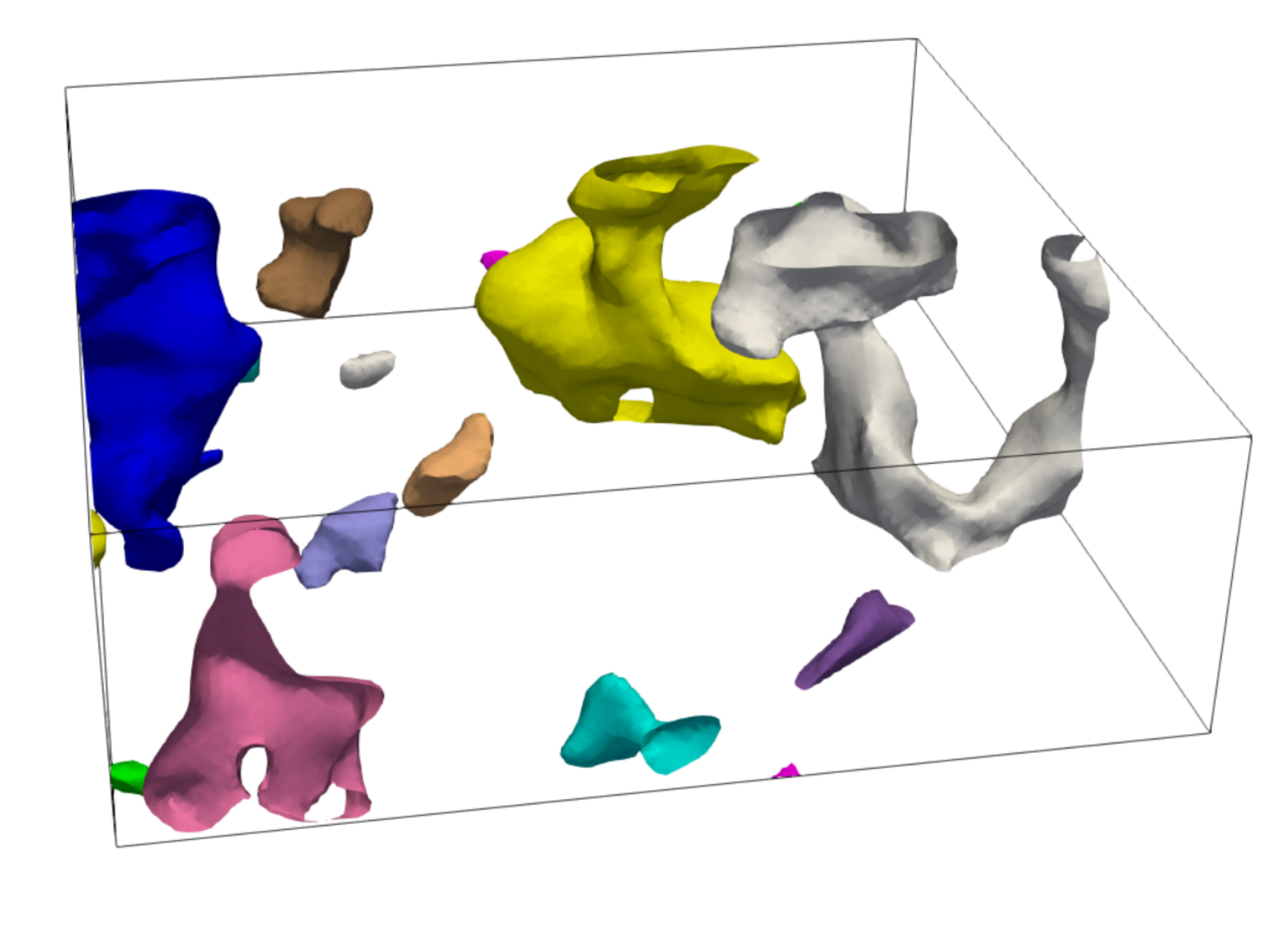}
         \caption{Cloud triggering features.}
     \end{subfigure}
     \hfill
     \begin{subfigure}[b]{0.23\textwidth}
         \includegraphics[width=\columnwidth]{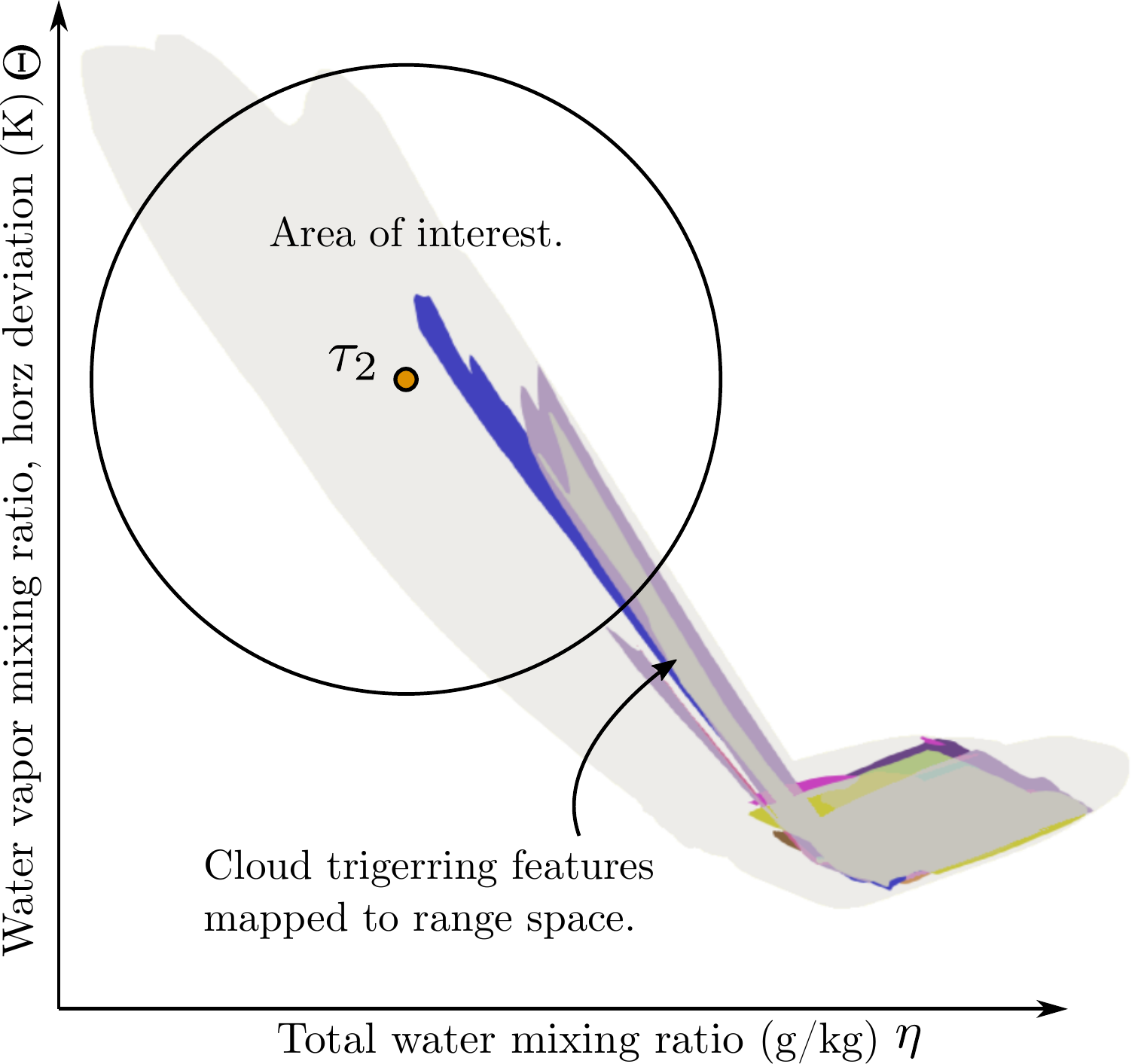}
         \caption{Mapped onto range space.}
     \end{subfigure}  
     \caption{Cloud triggering features in the atmosphere (a), defined as isosurfaces of a passive tracer concentration.
     Features mapped onto range space b) to understand their underlying behavior.
     The trait point is in an area interest for cloud formation.}
     \label{fig:clouds-intro}
\end{figure}

The third dataset comes from studying the processes that trigger convective clouds, which is important for large scale weather and climate predictions but is not fully understood.
One method used is to inject a passive tracer gas into numerical simulations to track bulk movements of air~\cite{denby2022,
couvreuxResolvedParametrizedBoundaryLayer2010}.
Certain concentrations of this tracer gas define cloud triggering features, which can be rendered as isosurfaces (see \cref{fig:clouds-intro}).
Studying the physical properties of these features, especially in terms of humidity $\eta$ and temperature $\theta$ is important for more accurate modeling.
Feature level sets have been successfully applied to this task with point traits whose feature level sets capture parts of features that contribute to cloud formation~\cite{Lei2025TraitDesign}.

\subsection{Comparative Evaluation}

In \cref{fig:toy-evaluation} we show the differences between the exact and approximate TIMTs for the three tetrahedra toy data set.
We have chosen four point traits outside $f(|M|)$ (the image of the domain in the range) and four traits inside.
The exact TIMT of any point trait chosen above the line defined by $f(ab)$ has a Y shape and below that, a single arc.
Differences between exact and approximate TIMT arise when the trait-induced graph introduces new nodes at edge minima (see \cref{fig:toy-example}), as can be seen for traits $\tau_4, \tau_5, \tau_7$ and $\tau_8$.

\begin{figure*}[p]
     \centering
     \begin{subfigure}[b]{0.22\textwidth}
         \includegraphics[width=\columnwidth]{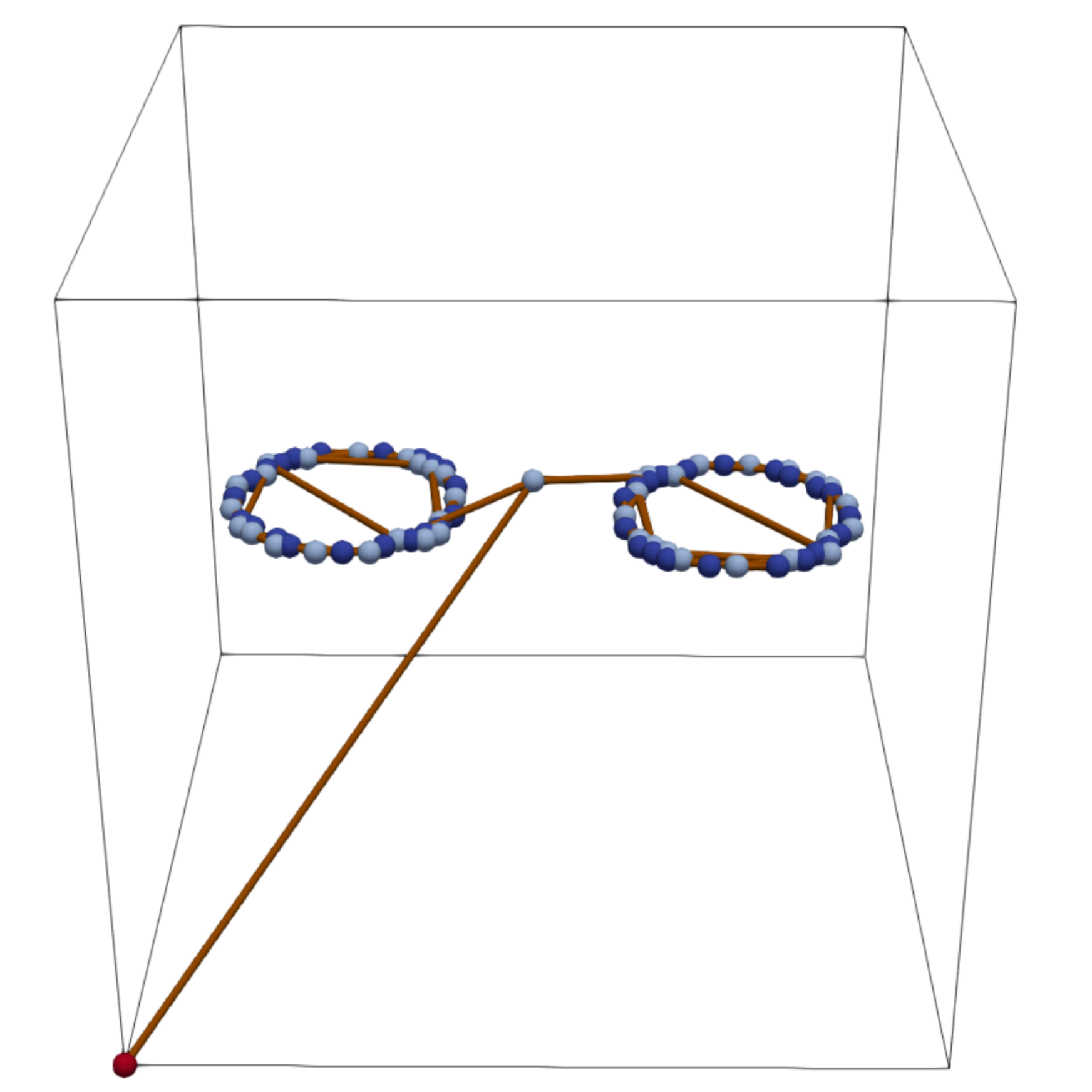}
         \caption{Approximate TIMT of $\tau_1$.}
     \end{subfigure}
     \hfill
     \begin{subfigure}[b]{0.22\textwidth}
         \includegraphics[width=\columnwidth]{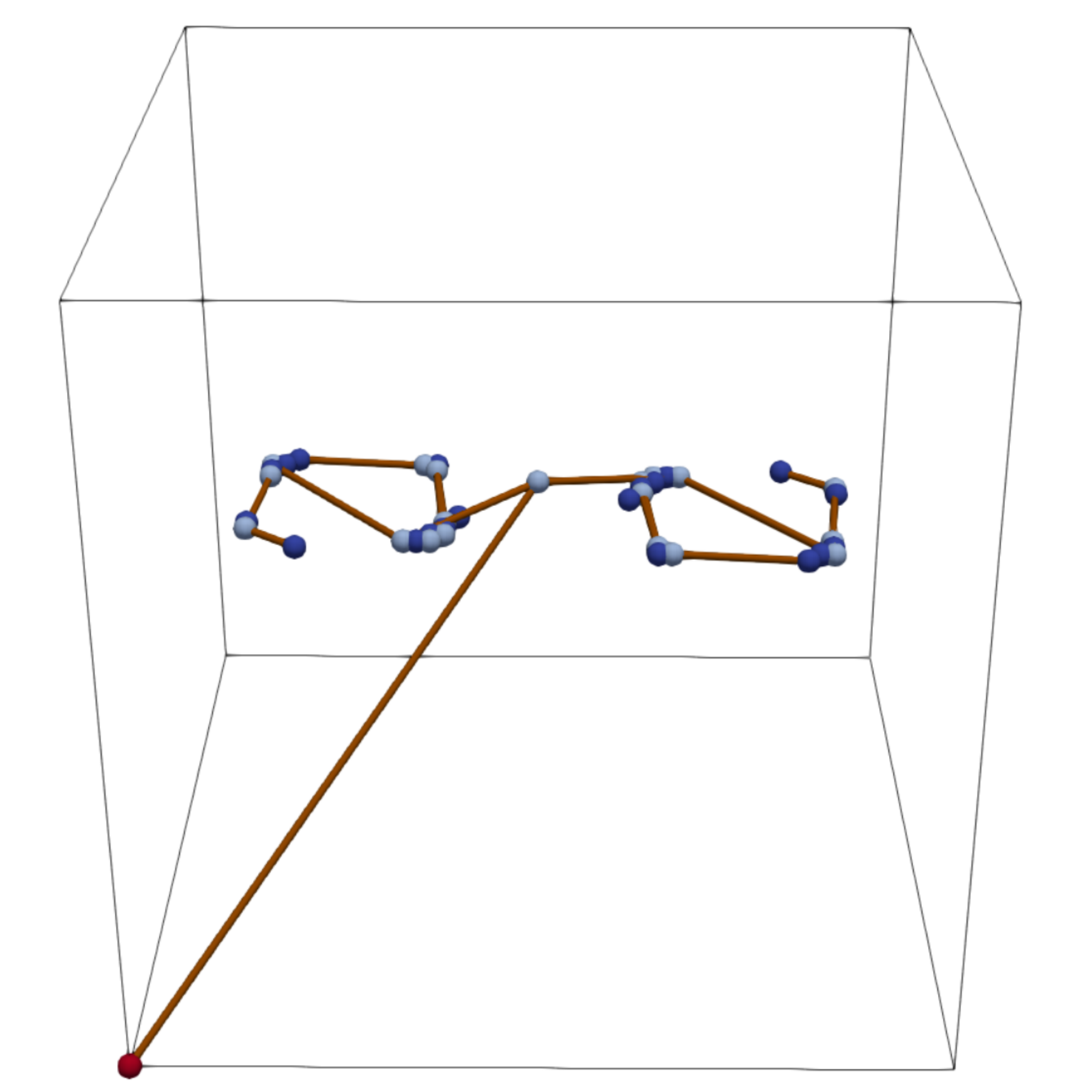}
         \caption{Simplified by $0.00002$.}
     \end{subfigure}
     \hfill
     \begin{subfigure}[b]{0.22\textwidth}
         \includegraphics[width=\columnwidth]{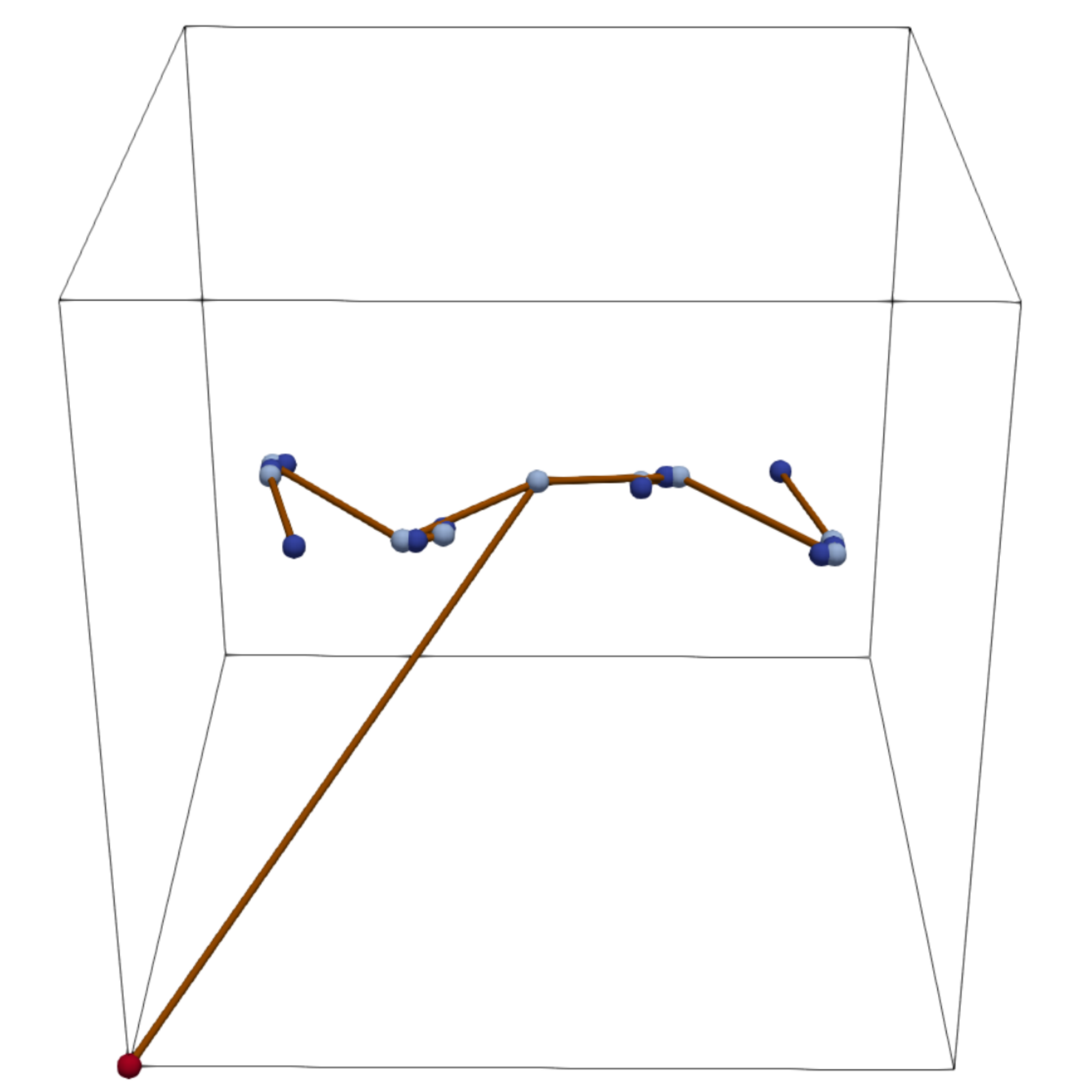}
         \caption{Simplified by $0.00003$.}
     \end{subfigure}
     \hfill
     \begin{subfigure}[b]{0.22\textwidth}
         \includegraphics[width=\columnwidth]{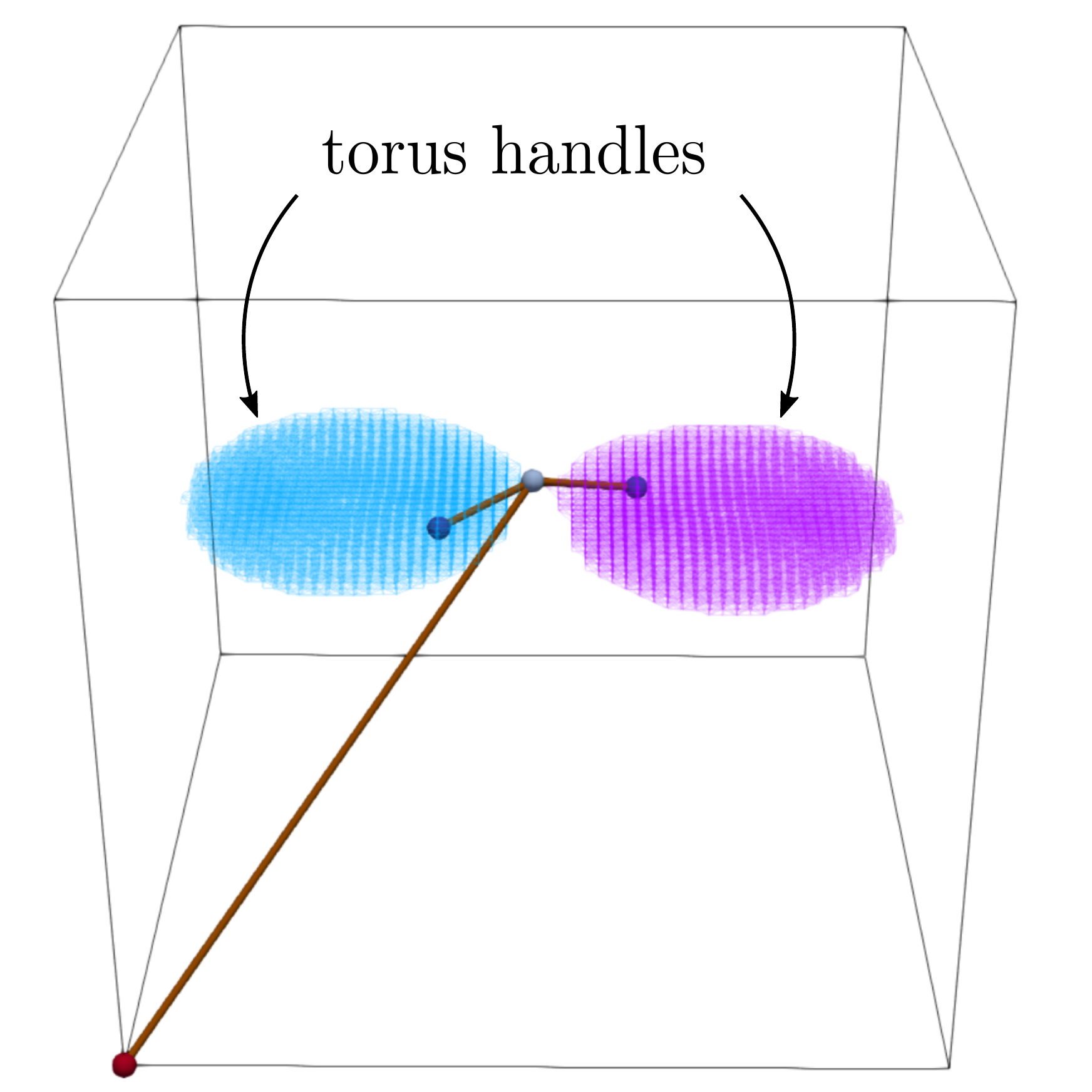}
         \caption{Simplified by $0.00004$ equals exact.}
     \end{subfigure}     
     \caption{Differences between the approximate and exact TIMTs of $\tau_1$ (see \cref{fig:torus-dataset}) at varying levels of persistence simplification embedded in the domain.
     The approximate TIMT (a) contains noise because $\tau_1$ is in the image of the input data in the range and so the zero fiber is not captured correctly.     
     At various levels of simplification, (b) and (c), the approximate TIMT approaches the exact one, until they become isomorphic (c) and both capture one of the main features of this dataset -- the two handles of the torus.
     }
     \label{fig:torus-evaluation}
\end{figure*}

\begin{figure*}[p]
    \centering
    \includegraphics[width=1\linewidth]{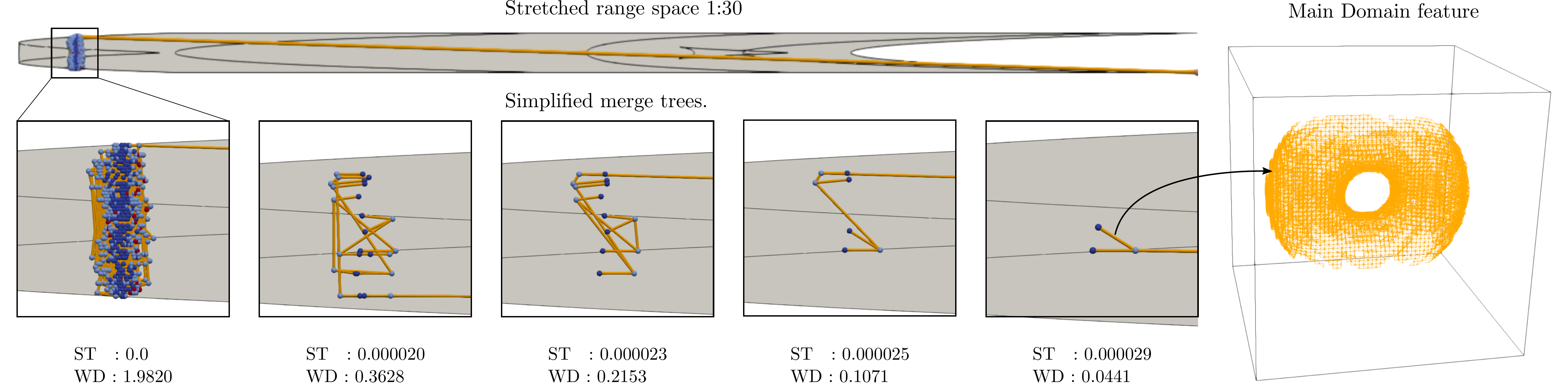}
    \caption{Differences between exact and approximate TIMTs for the trait $\tau_1$ in the torus example with the torus functions $f_1$ stretched 30 times with the TIMT embedded in the range space.
    This stretches the images of edges in the ranges, thus increasing the error of the approximate TIMT.
    Even though with simplification the distance between the exact and approximate trees becomes significantly smaller, the segmentation of the domain with respect to the approximate TIMT no longer correctly captures the two handles of the torus.
    ST denotes the simplification threshold, and WD denotes the Wasserstein distance between the exact and approximate TIMTs. The bottleneck distance between the exact and approximate TIMTs in all cases is 0.01327.
    }
    \label{fig:torus-differences}
\end{figure*}

\begin{figure}
    \includegraphics[width=\linewidth]{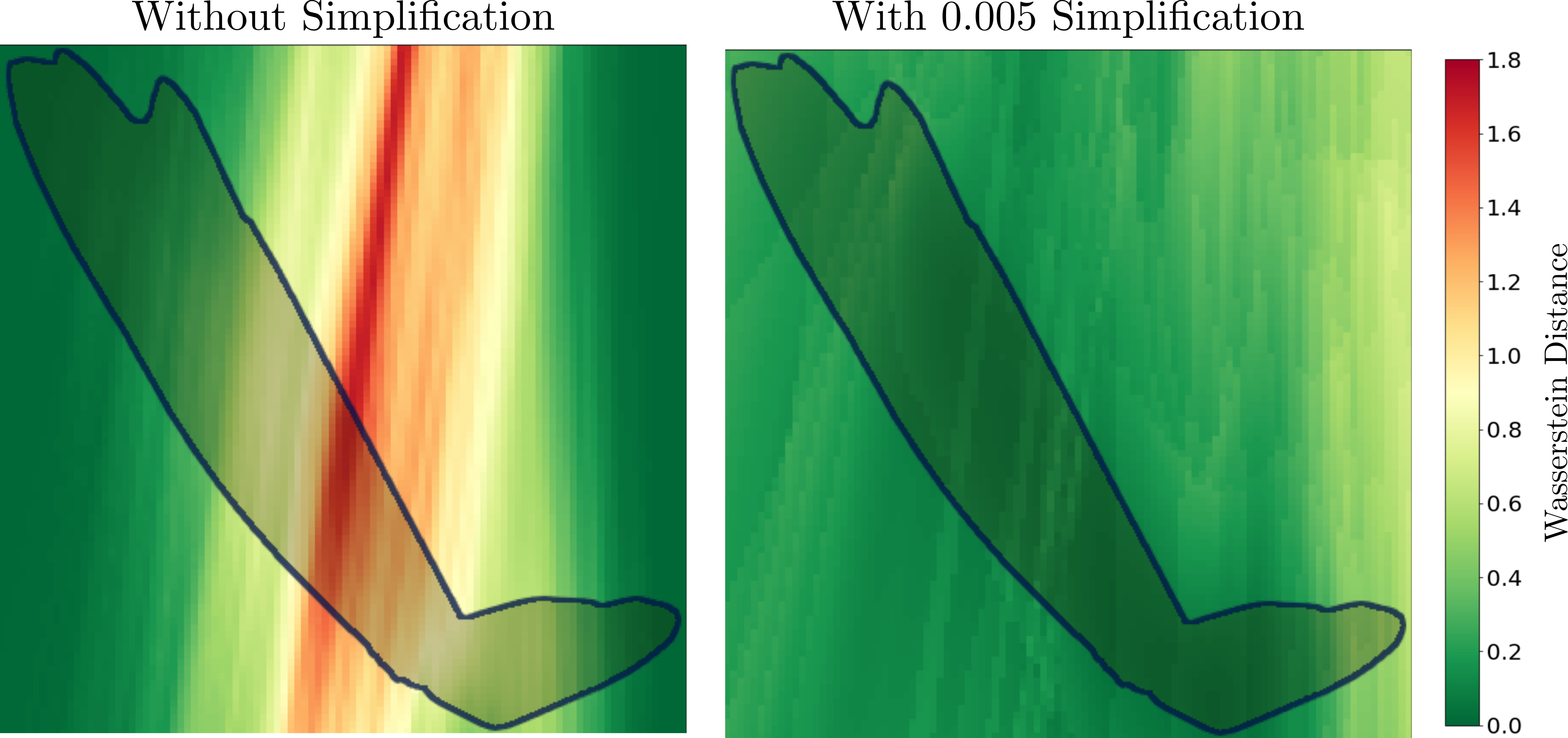}
    \caption{Distance matrix of the Wasserstein distance between the persistence diagrams of approximate and exact TIMTs overlaid on top of the range space.
    Trait points are sampled on a regular $100\times100$ grid over the min/max values of the two scalar fields $\eta$ and $\theta$.
    Without simplification (left) there can be significant differences between the two TIMTs, especially along the red line in the center.
    Simplification drastically reduces those quantitative differences.
    }
    \label{fig:clouds2-distance-matrix}
\end{figure}

\begin{figure}
    \centering    
    \includegraphics[width=1\columnwidth]{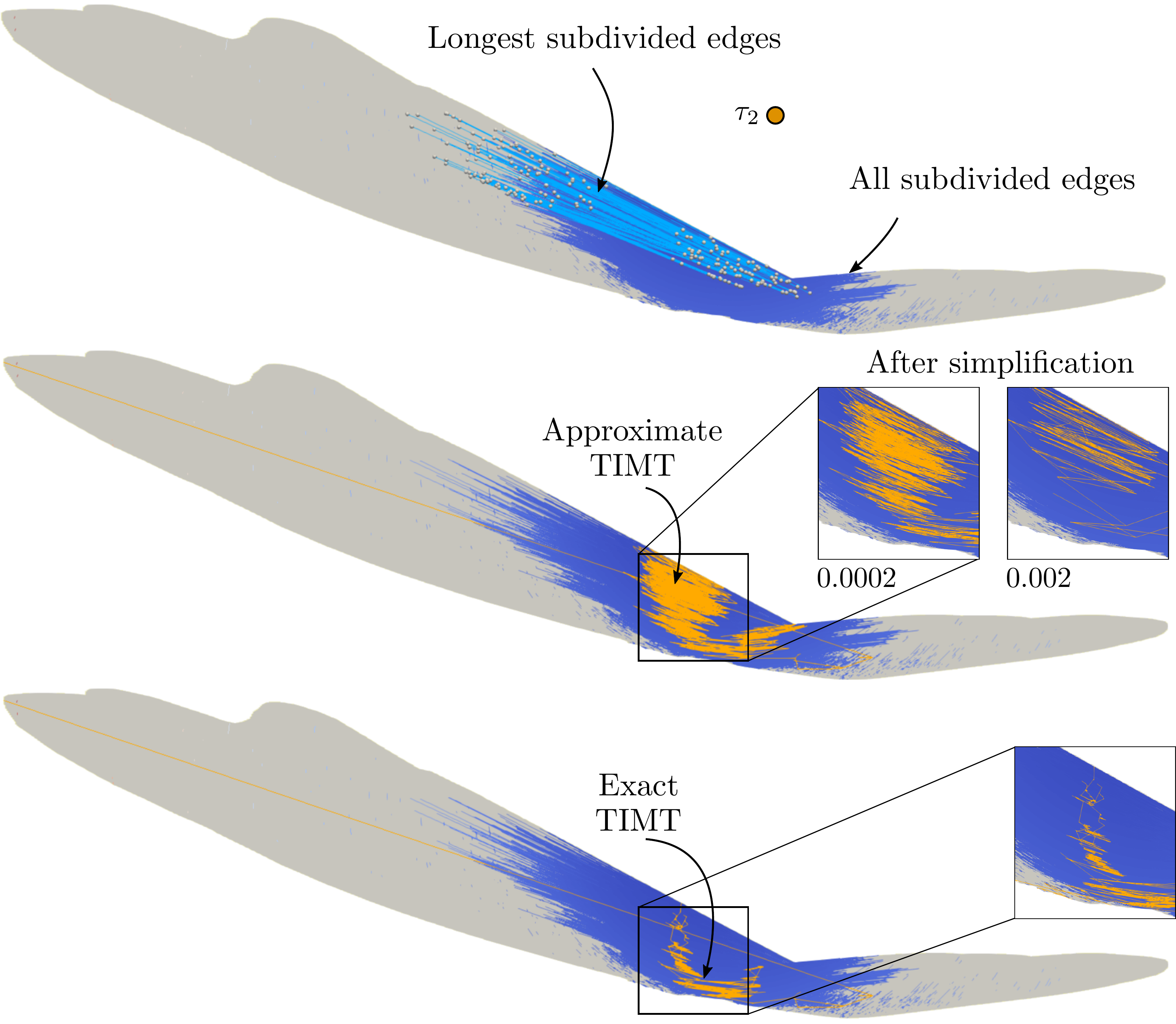}
    \caption{Qualitative comparison between the approximate and exact TIMTs for a trait point along the red strip of maximum Wasserstein distance from \cref{fig:clouds2-distance-matrix}.
    Differences arise from missing edge minima along edges which map to a long segments in the range (top) so the approximate TIMT (middle) contains significantly more noise than the exact one.
    This noise has high-persistence and simplification can remove potential features first.    
    }
    \label{fig:clouds2-qualitative}
\end{figure}

For the torus example, we selected a trait point whose fiber components are on the handles of the torus with the expectation to capture those features in the TIMT.
As expected, the approximate TIMT contains more noise than the exact one (see \cref{fig:torus-evaluation}).
However, after standard persistence based simplification, the approximate TIMT is structurally the same as the exact one, and they both capture the handles of the torus as the main features in their domain segmentation (see \cref{fig:torus-evaluation}(d)).

However, persistence simplification does not guarantee that the approximate TIMTs, or their induced segmentation, recover the exact results. 
\cref{fig:torus-differences} shows such a case, where the torus example is modified by replacing $f_1$ with $\bar f_1(x)=30\cdot f_1(x)$.
This stretches $f(|M|)$ along one attribute axis, increasing the projected edge lengths in the range and therefore the interpolation error of the approximate TIMT (see \cref{subsec:exact-approx-error}).

This setup reflects a common issue in multifield analysis: different fields often represent different physical quantities, measured in different units, and must be normalized or weighted before comparison. Outliers or unbalanced scalings can effectively stretch one attribute axis, as simulated here. As shown in \cref{fig:torus-differences}, simplification can reduce the tree distance between the approximate and exact TIMTs, but the induced segmentation may still miss the main features of interest, i.e., the two handles of the torus. A better approximation would require reducing projected edge lengths, for example, through mesh refinement.

Finally, we discuss the convective cloud triggering dataset.
We sampled 10,000 point traits in range space in a regular $100 \times 100$ grid based on the min/max values of the $\eta$ and $\theta$ scalar fields.
For each point trait, we compute the approximate TIMT and the exact one, as well as the Wasserstein distance between their persistence diagrams.
We plot the distance matrix of all distances in \cref{fig:clouds2-distance-matrix}, normalized to a unit square.
The first thing of note here is the almost vertical red strip of point traits where the distance is the largest.
The second thing is that simplification greatly reduces that distance.

Two key questions remain -- what causes the red strip in the distance matrix in \cref{fig:clouds2-distance-matrix} and does the simplification also qualitatively make the approximate and exact tree more similar?
We answer these in \cref{fig:clouds2-qualitative}.
In the top subfigure, we show the $f(|M|)$ in gray and all the edges that were subdivided (that with an edge minimum) in dark blue.
Note that this is not normalized and that the extent of the $\eta$ parameter is more than three times greater than the extent of the $\theta$ parameter.
In light blue, we show only the longest subdivided edges (at least half of the maximum edge length).
Similarly to the  torus dataset, here the range space is actually not square but stretched, so many edges become long segments in the range.
This causes the approximate TIMT to miss out on a lot of the structure of the trait-induced graph and to have level sets with a lot of erroneous discontinuities, leading to a lot of noise in the merge tree.

In the middle and bottom parts of \cref{fig:clouds2-distance-matrix} we visualize the approximate (middle) and exact (bottom) TIMTs in orange.
Due to our range embedding, we can immediately see that the extra noise in the approximate TIMT is actually due to those missed edge subdivisions.
Furthermore, as we simplify the approximate TIMT, we do not reach the structure of the exact one.
The noise in the approximate TIMT has high persistence and  gets simplified after the correctly captured combinatorial structure below it.
Again, like in the torus example, we have no guarantees that any large scale features would not be simplified.

\subsection{Performance}

In terms of performance, creating the trait-induced graph $G^\tau$ took up to 1 second for these datasets, up to 10 times longer than just computing the distance for the 1-skeleton.
This is to be expected because when computing the trait-induced graphs we have to determine the actual distance to every edge, which involved more geometric predicates and computation, which is done exactly.

In ~\cref{table:performance} we show the difference in the sizes of the 1-skeleton of the torus and convective clouds data sets for the traits $\tau_1$ and $\tau_2$.
While the trait-induced graph will always have at least as many vertices as the 1-skeleton, it typically has fewer edges.

\begin{table}[h]
\centering
\begin{tabular}{lcccccc}
\toprule
& \multicolumn{2}{c}{1-skeleton} & \multicolumn{2}{c}{$G^\tau$} & \multicolumn{2}{c}{Ratio (\%)} \\
\cmidrule(lr){2-3}\cmidrule(lr){4-5}\cmidrule(lr){6-7}
& V & E & V & E & V & E \\
\midrule
$\tau_1$ & 216K & 1,263K & 218K & 1,051K & 101\% & 83\% \\
$\tau_2$ & 320K & 1,871K & 393K & 1,777K & 122\% & 94\% \\
\bottomrule
\end{tabular}
\caption{Size differences between the 1-skeleton and the trait-induced graph for the point traits we used in our evaluation.}
\label{table:performance}
\end{table}

\section{Discussion and Conclusion}
\label{sec:discussion-conclusion}

We presented an exact graph-based construction for computing trait-induced merge trees of point traits over PL bivariate fields. The central observation is that, although the distance-to-trait function is not generally PL on the input tetrahedral mesh, its restricted sublevel sets inside each simplex are convex and therefore connected. This makes it possible to avoid a full subdivision into monotone regions: each tetrahedron contributes a single local monotone chain, and global connectivity is recovered by gluing these chains through the boundary events represented by mesh vertices, positive-valued edge minima, and connected components of the zero fiber.

We proved that the resulting trait-induced graph has the same sublevel-set connectivity as the exact distance field and therefore yields a merge tree isomorphic to the exact TIMT. We also related nonzero edge-based TIMT events to Jacobi edges of the underlying bivariate map, linking the critical behavior of the distance-to-trait field to the singular structure of the multifield. This relationship helps clarify where nonzero TIMT events may occur, while the full graph construction retains all edge-minimum events needed to represent sublevel connectivity
correctly.

Our comparison with the vertex-sampled approximation illustrates the practical consequences of ignoring within-cell variation of the distance field. The approximation overestimates the exact point-trait distance field, and its error is bounded by the maximum length of projected mesh edges in the range. This bound explains why the approximation improves under refinement, but also why it may fail when the trait lies in the range image or when projected edges are long. The experiments show that persistence simplification can sometimes reduce the apparent discrepancy, but it does not guarantee recovery of the exact TIMT or its induced
segmentation.

The present work focuses on point traits for bivariate maps on tetrahedral meshes. Several extensions are natural. For bivariate maps on domains of other dimensions, the simplex-wise convexity argument remains valid, although the dimension of inverse image $f^{-1}(q)$ changes and must be represented accordingly. For maps to higher-dimensional attribute spaces, the same convexity principle applies, but minimum events may occur on higher-dimensional faces of the image polytope, making boundary-event detection more involved. Traits such as lines, line segments, and finite point sets can also be approached by decomposing the attribute space into regions where the distance function reduces to simpler cases; a complete treatment requires additional interface events and a rigorous correctness proof.

Future work includes formalizing these extensions, improving scalability of the implementation, and investigating adaptive refinement driven by projected edge lengths in the range. We also plan to explore the relationship between TIMTs and Reeb spaces. Since TIMTs summarize the sublevel-set topology of distance functions induced by traits in attribute space, while Reeb spaces describe the connectivity structure of the underlying multifield map itself, understanding their interaction may lead to a more unified view of trait-based and inverse-image-based multifield topology.

\acknowledgments{This work is partially supported by the Wallenberg AI, Autonomous Systems and Software Program (WASP), funded by the Knut and Alice Wallenberg Foundation; the Swedish Research Council (VR) under Grant 2023-04806; the Swedish e-Science Research Centre (SeRC); and the ELLIIT environment for strategic research in Sweden. We thank Leif Denby (University of Leeds, the U.K.) for providing the convective cloud dataset used in this work.}

\bibliographystyle{abbrv-doi-hyperref-narrow}

\bibliography{template}

\appendix
\onecolumn
\section{Detailed Algorithm for Constructing Exact Trait-induced Graph}

The following pseudo code is a detailed version of Algorithm~1 provided in the paper. The implementation is publicly available at \href{https://github.com/tbmasood/Exact_TIMT}{\texttt{https://github.com/tbmasood/Exact\_TIMT}}.

\begin{algorithm}[!h]
\caption{Detailed construction of the trait graph $G^\tau$ 
}
\label{alg:trait-graph-detailed}
\begin{algorithmic}[1]

\Require PL bivariate field $f : M \to \mathbb{R}^2$ and a trait point $q \in \mathbb{R}^2$
\Ensure Graph $G^\tau = (V,E)$ with node weights $g(x)=\|f(x)-q\|$

\Comment{\textbf{Step 1: Initialize graph nodes from mesh vertices}}

\State $V \gets \emptyset$, \quad $E \gets \emptyset$

\ForAll{vertices $v$ of $M$}
    \State Add node $n_v$ to $V$ with weight $g(v)$
\EndFor

\vspace{0.4em}
\Comment{\textbf{Step 2: Insert interior edge minima}}

\ForAll{edges $e$ of $M$ \textbf{in parallel}}
    \If{$g|_e$ has a positive-valued interior minimum at $p_e$}
        \State Add node $n_e$ to $V$ with weight $g(p_e)$
        \State Record incidence map $\textsc{EdgeNode}[e] \gets n_e$
    \EndIf
\EndFor

\vspace{0.4em}
\Comment{\textbf{Step 3: Detect fiber-active faces}}

\ForAll{faces $F$ of $M$ \textbf{in parallel}}
    \If{$q \in f(F)$}
        \State Mark $F$ as fiber-active
    \EndIf
\EndFor

\vspace{0.4em}
\If{$\exists$ fiber-active face}

\vspace{0.4em}
\Comment{\textbf{Step 4: Compute local zero-fiber pieces}}

\ForAll{tetrahedra $T$ of $M$ \textbf{in parallel}}
    \If{$T$ has fiber-active incident faces}
        \State Mark $T$ as fiber-active
    \EndIf
\EndFor

\vspace{0.4em}
\Comment{\textbf{Step 5: Assemble global zero-fiber components}}

\State Build auxiliary graph $H_q$:
\Statex \qquad Each node of $H_q$ corresponds to a fiber-active tetrahedron
\Statex \qquad Connect two nodes $T_i,T_j$ in $H_q$ iff
$T_i$ and $T_j$ share a fiber-active face $F$

\State Compute connected components of $H_q$

\ForAll{connected components $C$ of $H_q$}
    \State Add node $n_C$ to $V$ with weight $0$
    \ForAll{$T \in C$}
        \State Record incidence map $\textsc{FiberNode}[T] \gets n_C$
    \EndFor
\EndFor
\EndIf

\vspace{0.4em}
\Comment{\textbf{Step 6: Local connectivity}}

\ForAll{tetrahedra $T$ of $M$ \textbf{in parallel}}

    \State Initialize local node set $V_T \gets \emptyset$

    \State Add the four vertex nodes of $T$ to $V_T$

    \ForAll{edges $e$ of $T$}
        \If{$\textsc{EdgeNode}[e]$ exists}
            \State Add $\textsc{EdgeNode}[e]$ to $V_T$
        \EndIf
    \EndFor

    \If{$\textsc{FiberNode}[T]$ exists}
        \State Add $\textsc{FiberNode}[T]$ to $V_T$
    \EndIf

    \State Sort nodes of $V_T$ in nondecreasing order of their weights

    \For{$i = 1$ to $|V_T|-1$}
        \State Add edge between $V_T[i]$ and $V_T[i+1]$ to $E$
    \EndFor

\EndFor


\State \Return $G^\tau=(V,E)$

\end{algorithmic}
\end{algorithm}

\end{document}